\documentclass[fleqn,usenatbib]{mnras}

\usepackage{newtxtext,newtxmath}

\usepackage[T1]{fontenc}

\DeclareRobustCommand{\VAN}[3]{#2}
\let\VANthebibliography\thebibliography
\def\thebibliography{\DeclareRobustCommand{\VAN}[3]{##3}\VANthebibliography}

\usepackage{graphicx}	
\usepackage{amsmath}	

\usepackage{lineno}

\usepackage{longtable}

\title[Detection of C$_3$ in Titan with VLT-ESPRESSO]{Detection of C$_3$ in Titan with VLT-ESPRESSO}

\author[R. Rianço-Silva et al.]{
R. Rianço-Silva,$^{1, 2, 3}$\thanks{E-mail: rafael.silva@ucl.ac.uk (RRS)}
P. Machado,$^{2, 3}$
P. Rannou,$^{4}$, J. Martins$^{5}$, A.E. Lynas-Gray$^{1, 6, 7}$, G. Tinetti$^{8}$
\\
$^{1}$Department of Physics and Astronomy, University College London, Gower Street, WC1E 6BT London, United Kingdom\\
$^{2}$Instituto de Astrofísica e Ciências do Espaço, Universidade de Lisboa, OAL, Edifício Leste, Tapada da Ajuda, PT1349-018 Lisbon, Portugal\\
$^{3}$Departamento de Física, Faculdade de Ciências, Universidade de Lisboa, Edifício C8, Campo Grande, PT1749-016 Lisbon, Portugal\\
$^{4}$Université de Reims Champagne-Ardenne, Reims, France\\
$^{5}$Instituto de Astrofísica e Ciências do Espaço, Universidade do Porto, CAUP, Rua das Estrelas, PT4150-762 Porto, Portugal \\
$^{6}$Department of Physics, University of Oxford, Keble Road, Oxford OX1 3RH, United Kingdom \\
$^{7}$Department of Physics and Astronomy, University of the Western Cape, Bellville 7535, South Africa \\
$^{8}$NMES Faculty, King’s College London, Strand Building, Strand, WC2R 2LS London, United Kingdom
}

\date{Accepted XXX. Received YYY; in original form ZZZ}

\pubyear{2026}

\begin{document}
\label{firstpage}
\pagerange{\pageref{firstpage}--\pageref{lastpage}}
\maketitle

\begin{abstract}

Titan is regarded as a Solar System natural laboratory for studying atmospheric photochemistry and abiotic production of organic molecules on cold small exoplanets. Since the Cassini-Huygens mission ended, telescope observations enabled new detections of increasingly complex carbon-based molecules at infrared and sub-mm wavelengths, while the optical regime has been largely overlooked. Following a recent tentative detection of the 4050\AA \hspace{0.1mm} absorption band of C$_3$ in Titan – a photochemical precursor to aromatic chemistry – in archived optical VLT-UVES spectra (R $\approx$ 60 000), this work presents a 8$\sigma$ detection of the C$_3$ 4050\AA \hspace{0.1mm} absorption band in Titan from dedicated ultra-high-resolution VLT-ESPRESSO observations of Titan (R $\approx$ 190 000, highest spectral resolution observations of Titan in optical wavelengths ever). VLT-ESPRESSO spectrum is compared to a model spectrum of Titan, for varying C$_3$ abundances; a $\chi^2$ curve is drawn to assess the agreement of non-solar spectral features with C$_3$ absorption when varying C$_3$ abundance; and a Bayesian Monte-Carlo-Markov-Chain (MCMC) fit between model and observed spectra is performed. $\chi^2$ curve analysis yields an 8$\sigma$ C$_3$ detection, consistent with a C$_3$ column density of $N = 1.5 \times 10^{13}$ cm$^{-2}$, whereas the MCMC fit retrieved a C$_3$ column density on Titan’s atmosphere of $N = (1.47 \pm 0.30) \times 10^{13}$ cm$^{-2}$ at 5$\sigma$, the same order of magnitude of predicted C$_3$ abundances by photochemical models, reaching ppm levels on Titan’s mesosphere. This work showcases the usefulness of instruments and techniques originally dedicated to exoplanet research when applied to solar system targets and science cases.

\end{abstract}




\section{Introduction}

\subsection{On Titan's atmosphere}

\par Titan, with its dense and chemically rich atmosphere and likely subglacial water ocean, stands out amongst the icy moons as one of the most extensively studied bodies in the solar system \citep{Horst2017} \citep{MacKenzie2021}. The diversity and uniqueness of Titan's distinct environmental physical and chemical characteristics have sparked a long-lasting interest in the largest satellite of Saturn, covering a wide array of scientific areas, from atmospheric chemistry to planetary geology and even to astrobiology \citealt{Horst2017, MacKenzie2021}.

\par A major reason for Titan's uniqueness is the unusually large and dense atmosphere for an icy satellite, reaching Earth-like densities and pressures on its troposphere \citep{SanchezLavega2011} and sharing the same major atmospheric constituent (N$_2$) with the Earth's atmosphere \citep{Horst2017}. The photochemical dissociation of atmospheric CH$_4$ (and to a lesser extent N$_2$) on Titan's upper atmosphere leads to a complex chain of photochemical reactions that give rise to a diverse set of increasingly complex hydrocarbons, nitriles and organic compounds \citep{MacKenzie2021}. The vast photochemical production of organic compounds on Titan creates precursors to the organic hazes that enshroud its the atmosphere, a process akin to the one that is believed to have occurred on the early Earth \citep{Trainer2006}. While the extent of its contribution to the emergence of life on Earth is unclear, it is believed that the production of photochemical organic hazes could take place across the Galaxy on Titan-like exoplanets \citep{Arney2017}.

\par Understanding the chemical complexity of Titan's atmosphere has benefited from parallel contributions from observations and modelling work \citep{Horst2017}. Early remote sensing observations from Voyager 1 and ground-based telescopes have hinted at the chemical complexity of Titan's atmosphere which motivated preliminary studies on the atmospheric production of organic molecules \citep{Yung1984}. The Cassini-Huygens space mission enabled an in-depth study of Titan's photochemical processes and products, characterizing profiles of higher-order hydrocarbons and nitriles (e.g. C$_2$N$_2$, HCN, HC$_3$N, C$_4$H$_2$ and C$_3$H$_4$,  \cite{Teanby2009}, \cite{Coustenis2010}) as well as simple aromatic compounds like benzene (C$_6$H$_6$) \citep{Vuitton2008}.

\par Since the end of the Cassini mission in 2017, the observational study of Titan's atmosphere has been carried forward with several new molecular discoveries from ground-based sub-mm (e.g., detection of c-C$_3$H$_2$ with ALMA, \cite{Nixon2020}) and infrared high-resolution observations (e.g. detection of CH$_2$CCH$_2$ with IRTF-TEXES \cite{Lombardo2019}). Recent space-based observations from near to mid-infrared observations with JWST have also contributed to this effort, directly detecting the methyl radical, CH$_3$, one of the main actors on Titan's atmospheric photochemistry \citep{Nixon2025}. 
\par It is worth noticing that despite being less challenging than infrared observations due to the transparency of Earth's atmosphere, ground-based high-resolution observations of Titan in visible wavelengths have been largely overlooked \citep{Rianço-Silva2024}. One of the reasons behind this is the fact that most of the strongest molecular absorption features of the molecules present in Titan's atmosphere are present in the near and mid infrared spectral regions \citep{Nixon2025}. A notable exception to this is precisely the object of study of this work - the molecule of propadienediylidene (C$_3$) also known as tricarbon \citep{Lynas-Gray2024}. 

\subsection{On the molecule of C$_3$}

\par Described for the first time in an astrophysical setting by \cite{Douglas1951}, on the active coma of a comet, propadienediylidene (or tricarbon, from here on, C$_3$) has ever since been found across a wide range of carbon-rich, energetic environments - in particular, interstellar dark clouds \citep{Schmidt2014} and the atmospheres of carbon-rich stars \citep{Swings1953}. These early astrophysical detections of C$_3$ relied on its strong spectral features on the violet section of the visible spectrum at 4050\AA \hspace{0.2mm} which \cite{Gausset1965} identified as the $\tilde{A}^1 \Pi_u - \tilde{X}^1\Sigma^+_g$ 000-000 electronic transition. With the advent of very high-resolution spectroscopy, a continuous work of improvement of the resolution of molecular linelists of C$_3$ \citep{Tanabashi2005} has allowed the detection of C$_3$ 4050\AA \hspace{0.2mm} absorption on the interstellar medium with visible high resolution spectrographs such as VLT-UVES with resolving powers of R > 60 000 (e.g. \cite{Schmidt2014}, \cite{Fan2024}).

\par The thought of Titan's upper atmosphere as a carbon-rich energetic environment with complex photochemical networks has led modelling works to consider the effect of highly unsaturated C$_3$-hydrocarbons on their photochemical networks. Remarkably, \cite{Herbad2013} and \cite{Dobrijevic2016} predicted that C$_3$ could be among the most abundant 3 carbon species present in the upper atmosphere of Titan, reaching molar ratios of the order of ppm at 400 km (lower mesosphere) of tens of ppm at altitudes of 800 km (upper mesosphere) \citep{Herbad2013} - equivalent to a maximum predicted column density of C$_3$ in Titan's atmosphere of $5 \times10^{13}$ cm$^{-2}$. These relative abundances are comparable with those of the most abundant detected 3-carbon hydrocarbons, such as propane (C$_3$H$_8$) at the ppm level across Titan's stratosphere and mesosphere \citep{Dobrijevic2016}. 

\par Indeed and unlike more chemically stable 3-carbon hydrocarbon species (C$_3$H$_8$, C$_3$H$_6$, CH$_2$CCH$_2$ and CH$_3$CCH) which are observed in Titan's stratosphere \citep{Lombardo2019}, C$_3$ is predicted to be significantly depleted in Titan's stratosphere, with relative abundances plummeting below ppt levels at altitudes lower than 300 km \citep{Herbad2013}. This effect has also been described and observed for c-C$_3$H$_2$ (\cite{Nixon2020} and \cite{Willacy2022}), which is the most similar species to C$_3$ detected in Titan up to this point. Possible reasons for this distinct vertical profile peaking in relative abundance at the upper mesosphere and stratospheric depletion for both C$_3$ and c-C$_3$H$_2$ could be traced to the increase in the reaction with H atoms in the lower mesosphere and stratosphere, described in \citep{Dobrijevic2016} and \citep{Willacy2022}.

\par The C$_3$ molecule is particularly interesting for the evolution of chemical complexity in Titan's atmosphere, as it is thought to play a role on the production of aromatic chemistry in Titan \citep{Loison2019} since its reaction with the CH$_3$ radical significantly impacts the production of C$_4$H$_2$ \citep{Mabel2023}. C$_4$H$_2$ is then a direct precursor of benzene and other aromatics in Titan (including PAH) \citep{Loison2019} which are known to play a pivotal role in prebiotic processes \citep{Ehrenfreund2006}. Hence, given the great interest for astrobiology of the atmospheric production of  aromatic species, it is critical to gain understanding of the many "missing links" - such as the reaction rates of C$_3$ + CH$_3$ which are challenging to study experimentally \citep{Mabel2023}. Detecting and retrieving the atmospheric abundance of C$_3$ in Titan will therefore provide key observational constraints to the models that attempt to describe the production of prebiotic molecules on the atmosphere of Titan \citep{Loison2019}.

\par However, in spite of its importance and high predicted abundance in Titan's mesosphere, no direct and conclusive detection of C$_3$ has been reported thus far. Possible reasons for this might be connected to the lack of an infrared, detailed linelist of C$_3$ - recently addressed by the ExoMol project \citep{Lynas-Gray2024} - and the largely overlooked visible spectrum of Titan, where the strongest spectrum features of C$_3$ are present \citep{Rianço-Silva2024}. The sole indirect evidence for the presence of C$_3$ in Titan stems from the INMS peak at 37 m/z units attributed to the C$_3$H$^+$ \citep{Vuitton2007} which is the most likely loss mechanism for C$_3$ \citep{Dobrijevic2016}.

\par Apart from this, a recent study with VLT-UVES has claimed a tentative detection of the C$_3$ 4050\AA-band in Titan's atmosphere from very high resolution visible spectroscopy \citep{Rianço-Silva2024}. Despite this encouraging result, which was consistent with the presence of C$_3$ in Titan at a column density of $10^{13}$ cm$^{-2}$, the small number and the low SNR of detected spectral features of C$_3$ in \cite{Rianço-Silva2024} require further observations at increased spectral resolution and SNR to allow confirming the detection of C$_3$ in Titan. This is the purpose of this work which is organized as follows: In section \ref{sec:Observations} we describe the VLT-ESPRESSO observations of Titan and their data reduction. In section \ref{sec:Model} we describe our model for Titan's spectrum with C$_3$ absorption. In section \ref{sec:results} we show our observational results and compare them to the spectral model of Titan through a $\chi^2$ analysis and through a MCMC Bayesian retrieval and in section \ref{sec:Conclusions} we discuss and conclude.

\section{Observations and Data Reduction}
\label{sec:Observations}

\subsection{Observations}

The dedicated observations used for this work were obtained with the European Southern Observatory's Very Large Telescope (VLT) ESPRESSO (Échelle SPectrograph for Rocky Exoplanets and Stable Spectroscopy Observations) instrument. VLT-ESPRESSO is a visible, High Resolution, fiber-fed, cross-dispersed, échelle spectrograph at the Incoherent Combined-Coudé Laboratory (ICCL) of the VLT, allowing this instrument to receive light from any of the four 8m-class Unit Telescopes \citep{Pepe2014}. Although originally built with the goal of searching for and characterizing exoplanets, VLT-ESPRESSO has already proven to be a very useful instrument for solar system science, having enabled Doppler velocimetry studies of the atmosphere of Jupiter \citep{Machado2023}.

\par Titan was observed for 2h20m with VLT-ESPRESSO, starting at 00:17 UTC of 4 December 2024, as part of the ESO observing program 114.277N, using the Ultra-High-Resolution (UHR) mode of VLT-ESPRESSO \citep{Pepe2021}. This is the highest spectral resolution mode available with VLT-ESPRESSO, reaching a resolving power of 190 000 - making these the observations of Titan at the highest spectral resolution ever done in optical wavelengths. During the course of the observation the seeing (FWHM) was below 0.8 arcsec and the sky was clear. Titan's solid body (Diameter = 5150km) corresponded on the observation night to an apparent angular diameter of 0.74 arcsec. However, due to its thick, opaque atmosphere, Titan's optical radius at $\lambda = $ 400nm is about 300km high in its atmosphere \citep{Lorenz1999}. This yields a Titan optical diameter of 5750km at $\lambda = $ 400nm, corresponding to an apparent angular diameter of 0.826 arcsec on the observation night. This implies that the VLT-ESPRESSO 0.5 arcsec fiber was fully covered by Titan's disk - with implications for modelling Titan's atmosphere described on appendix \ref{Appendix}. At the time, Titan had an apparent V-band magnitude of +8.62, a surface brightness of 7.71 mag/arcsec$^2$ and an illuminated fraction of 99.73\%. Target ephemerides were obtained from the Horizons System - Solar system Dynamics Ephemerides Calculator from NASA/JPL \citep{Horiz}.

\par The set of 7 exposures (each 20 minutes long) of Titan enabled a coverage of the entire visible, backscattered spectrum of Titan (at the wavelength range of 378.2–788.7 nm), reaching a Signal to Noise Ratio (SNR) of 60 at 405 nm on each exposure. This would imply a total SNR close to 100, after stacking the 7 exposures.

\subsection{Data Reduction}
\subsubsection{Removing Telluric effects and stacking exposures}

\par The data were reduced with ESPRESSO Data Reduction Software version 3.3.0. After obtaining individual flux-calibrated spectra for each Titan exposure, we removed the effects of the variation of air mass in multiple exposures, which are evident in figure \ref{fig:7_spectral_exposures}. To do so, we obtained a third-degree polynomial fit of the division of each exposure's Titan spectrum by the spectrum of the exposure at the lowest airmass (see fig. \ref{fig:7_spectral_exposures_polyfit}). The spectrum of each exposure was then divided by the respective normalized polynomial fit to remove the effect of varying air mass on the spectral continuum (a value between 0 and 1 for each spectral wavelength bin). The spectral flux errors, $\delta S_{\lambda, i}$, associated to the $i$-th exposure spectral flux measurement, $S_{\lambda, i}$, were obtained in a similar fashion: each exposure's spectral flux error ($\delta S_{\lambda, i}$) was divided by the respective normalized polynomial fit to the airmass (a value between 0 and 1 for each wavelength bin), maintaining each exposure's SNR after removing the effect of varying airmass across exposures. After this, the Titan spectra (corrected for airmass variation, as shown in fig. \ref{fig:7_spectral_exposures_Airmass_corrected}) were averaged together to obtain a single Titan spectrum, $S_\lambda$, for the entire observation night (eq. \ref{spectral_average}).

\begin{equation}
    S_\lambda = \frac{1}{N} \sum^N_i S_{\lambda, i}
    \label{spectral_average}
\end{equation}

The respective spectral flux errors for each $i$ exposure were also combined into a single spectral flux error per wavelength bin, $\delta S_\lambda$, as shown in eq. \ref{spectral_error_average}, the equation for the standard error of the mean, which depends on 2 terms: the first term accounts for the average of the intrinsic spectral flux error ($\delta S_{\lambda, i}$) per wavelength bin, per exposure, i.e. the spectral flux error associated to each spectral flux measurement per each exposure. The second term depends on the scatter of the spectral flux values per exposure and wavelength bin, ($S_{\lambda, i}$) around their average, obtained in equation \ref{spectral_average}, corresponding to the variance scatter between the multiple stacked exposures ($s^2$). Both terms are summed in quadrature in equation \ref{spectral_error_average}, allowing us to account for both the intrinsic uncertainties in each exposure as well as the scatter between the different exposures. Stacking these spectra increased the total SNR, reaching $\approx$ 100 at 405 nm, as shown in figure \ref{fig:7_spectral_exposures_Airmass_corrected_average}.

\begin{equation}
    \delta S_\lambda = \sqrt{\frac{1}{N} \left(\frac{1}{N} \sum^N_{i=1} \delta S_{\lambda, i}^2 + s^2 \right)}, \text{ where } s^2 = \frac{1}{N-1}\sum^N_{i = 1}(S_{\lambda, i} - S_\lambda)^2
    \label{spectral_error_average}
\end{equation}

One extra step in data reduction involves normalizing the average Titan spectrum by its spectral continuum, to make it directly comparable with the normalized solar spectrum as described in section \ref{sec:Model}. To do so, we obtain the spectrum continuum using a Savitzky-Golay filter \citep{Savitzky-Golay} as shown in figure \ref{fig:spectrum_normmalization_SG}.

It is also worth noticing that the spectral flux errors per exposure are dominated by the photonic error, as expected from ESPRESSO's exposure time calculator (ETC) \citep{Pepe2021}, where for the conditions of these observations and for a target akin to Titan, the combined contributions to the error from sky, dark current and read-out-noise are below 2\% of the photonic noise from Titan's brightness. Hence, SNR roughly scales with the square root of the observation time, as these observations are photon-noise limited.

\par After removing the continuum telluric contributions to the observed spectrum, we solely focus on the spectral region of interest corresponding to the C$_3$ $\tilde{A}^1 \Pi_u - \tilde{X}^1\Sigma^+_g$ 000-000 band, on the narrow wavelength region between 4040\AA \hspace{0.5mm} and 4060\AA \hspace{0.5mm} as shown in figure \ref{fig:plot_67_errors}. This allows a search for possible telluric absorption or emission lines that could have contaminated the spectral section that is crucial for this analysis. Following the approach of \cite{Dias2022} for high spectral resolution observations of other solar system targets, we used NASA's open-access Planetary Spectrum Generator (PSG) radiative transfer suite to simulate the transmission of the Earth's atmosphere \citep{Villanueva2018} for VLT. From the simulated transmission spectrum of Earth's atmosphere at the ESPRESSO UHR mode resolution (shown in grey on Fig. \ref{fig:C3_sweeping_original}), no telluric absorption or emission line causing a variation of transmission above 0.01\%  is present in the 4040\AA \hspace{0.5mm} and 4060\AA \hspace{0.5mm} wavelength range, well bellow the sensitivity allowed by these observations high SNR, thus discarding the need for additional telluric correction.

\subsubsection{Removing Doppler shifts}

\par A key step to take into account when dealing with observations of solar system objects in high spectral resolution is the correction of the orbital Doppler shift. This Doppler shift is consequence of the relative motions between the observer (on Earth) and the observed body, causing a shift on the observed wavelengths $\Delta \lambda$ given by

\begin{equation}
        \Delta \lambda = \lambda_0 \frac{v}{c}
        \label{Doppler_shift}
\end{equation}

where $\lambda_0$ is the rest-frame wavelength, before Doppler shift occurs. The absorption lines that we aim to detect on Titan's atmospheric spectra ought to occur on Titan's rest frame. Hence, it is key to shift the observed spectrum to Titan's rest frame, by subtracting from each wavelength bin the associated Doppler shift, given by eq.\ref{Doppler_shift}. For reference, during the observation, Titan's radial velocity with respect to the observer at VLT was of 5.33 km/s, towards the observer (from \cite{Horiz}), which amounts to a Doppler shift of 0.072\AA \hspace{0.1mm} at 4050\AA.

\section{Modelling C$_3$ absorption in Titan}
\label{sec:Model}

When analysing the observed spectrum, it makes sense to directly compare it with the modelled spectrum of Titan in the wavelength regions of interest. Here we follow the approach used in \cite{Rianço-Silva2024}, whereby this very short wavelength span of Titan's spectrum (2 nm-wide) is modelled as a normalized backscattered solar spectrum upon which occurs the C$_3$ line absorption. This is a valid approximation since 

\begin{itemize}
    \item The absorption of C$_3$ is expected to occur on the upper layers of Titan's atmosphere (altitudes above 400 km), significantly above the optical radius of Titan at 405 nm (below 250 km, from \cite{Lorenz1999}) where backscatter occurs.

    \item The spectrum of Titan in this short wavelength section is dominated by continuum haze absorption and scatter, which do not produce high-resolution features. In \cite{Rianço-Silva2024}, it was shown that the continuum haze absorption of Titan varies by less than 0,04\% per \AA ngstrom over the 400 nm to 500 nm region, which is negligible for the wavelength range and SNR of the observed spectrum.

    \item There are no other known absorbing gases in Titan's atmosphere (apart from C$_3$) that showcase absorption lines on the 4040\AA \hspace{0.2mm} to 4060\AA \hspace{0.2mm} wavelength range, as was also shown in \cite{Rianço-Silva2024}.    
\end{itemize}

\par Hence, we model the UHR spectrum of Titan in the 4040\AA \hspace{0.2mm} to 4060\AA \hspace{0.2mm} wavelength range as a product of a UHR solar spectrum (the \cite{Kurucz2006} solar spectrum at R = 500 000, with a spectral resolution degraded to ESPRESSO UHR R = 190 000) multiplied by the modelled C$_3$ transmission spectrum, $M(\lambda, T, N)$. 

\par The the C$_3$ transmission spectrum, $M(\lambda, T, N)$, was based on the most recent linelist of C$_3$ covering its $\tilde{A}^1 \Pi_u - \tilde{X}^1\Sigma^+_g$ 000-000 band, by \cite{Fan2024}, by combining multiple previous linelists, such as the ones from \cite{Tanabashi2005} and \cite{Schmidt2014}. It is worth noticing that these linelists were obtained through comparison with observational data from VLT-UVES, with a spectral resolution of R $\approx$ 80 000 \citep{Dekker2000}, less than half of the UHR mode of VLT-ESPRESSO. These linelists solely contain the following information for each C$_3$ line: its wavelength, $\lambda_{j}$, its corresponding transition within the P, Q and R branches (associated to its J-level), and its oscillator strength, $f_{j}$ \citep{Fan2024}. 

\par The first step to model the C$_3$ transmission spectrum in Titan's atmosphere is to obtain its partition function across the multiple molecular energy states ($J$) associated with the vibronic transitions. Assuming thermal equilibrium, this partition is given by a Boltzmann distribution as shown in eq. \ref{eq:Boltzman}, where B = 0.431 cm$^{-1}$, from \cite{Tanabashi2005}. For the partition function sum at the denominator of eq.\ref{eq:Boltzman}, $k$ are summed across all the observed rotational levels, from \cite{Fan2024}.

\begin{equation}
  p(J, T) = \frac{(2J + 1)\exp{\left[-\frac{BJ(J+1)}{k_BT}\right]}}{\sum\limits_{k} \left( (2k + 1)\exp{\left[-\frac{Bk(k+1)}{k_BT}\right] }\right)}
    \label{eq:Boltzman}
\end{equation}

\par This distribution of the gas molecules across the multiple states (J) is used to calculate the line strength, $S_j$, of each transition from the wavelengths, $\lambda_{j}$, and oscillator strengths, $f_j$, provided in the C$_3$ linelists, as shown in eq. \ref{eq:Line_Strength} \citep{SanchezLavega2011}. Considering thermal equilibrium, the line profiles of individual absorption lines of C$_3$ are Gaussian and given by $G_j(\lambda, T)$, as described in eq. \ref{eq:Line_Profile}, where the Doppler broadening coefficient, $\gamma_{j,D}(T)$, is given by eq. \ref{Doppler_broadening}, which depend on $m_{C_3}$, the mass of the C$_3$ molecule.

\begin{equation}
    S_j = \frac{e^2\lambda_j^2}{4m_ec^2\varepsilon_0}p(J,T).f_j
    \label{eq:Line_Strength}
\end{equation}

\begin{equation}
    G_j(\lambda, T) = \frac{S_j}{\gamma_{j,D}(T).\sqrt{\pi}}.\exp \left[ - \left( \frac{\lambda - \lambda_j}{\gamma_{j,D}(T)}\right)^2\right]
    \label{eq:Line_Profile}
\end{equation}

\begin{equation}
    \gamma_{j,D}(T) = \frac{\lambda_j}{c}\sqrt{\frac{2k_BT}{m_{C_3}}}
    \label{Doppler_broadening}
\end{equation}

\par Finally, the optical depth profile of each absorption line, $\tau_j(\lambda, T, N)$ is given by the product of the line profile $G_j(\lambda, T)$ and the column density of C$_3$ in Titan's atmosphere, $N$ and a factor of $\tilde{A} \simeq$ 2.2155 which accounts for the spherical geometry of Titan's atmosphere in eq. \ref{eq:optical_depth}. This accounts for the varying slant paths of photons across the illuminated disk, by adopting an effective path length enhancement factor of $\tilde{A} \simeq$ 2.2155, integrating the effect of the varying airmass over the observed disk of Titan compared to the vertical two-way path at the disk centre). This factor is an approximation for a Lambertian sphere, where photons near the planetary limb traverse a longer atmospheric column \citep{Palmer2001}. 

\begin{equation}
    \tau_{j}(\lambda, T, N) \approx \tilde{A}G_j(\lambda, T)N
    \label{eq:optical_depth}
\end{equation}

\par In order to get the complete model transmission spectrum of the C$_3$ absorption band on Titan, $M(\lambda, T, N)$, we multiply all of the individual lines (of all absorption lines identified in \cite{Fan2024}) transmission profiles, $M_j(\lambda, T, N)$, which result from the Beer-Lambert law considering a 2-way path of light down to the optically thick layer where backscattering is assumed to occur (effective airmass of $\tilde{A} \simeq$ 2.2155, described above), as shown in eq. \ref{eq:Transmission}.

\begin{equation}
    M(\lambda, T, N) = \prod_j M_j(\lambda, T, N) = \prod_j e^{-\tau_j(\lambda, T, N)} \approx \prod_j e^{-\tilde{A}G_j(\lambda, T)N}
    \label{eq:Transmission}
\end{equation}

\par Finally, the synthetic spectrum of Titan is obtained by multiplying the Transmission spectrum of C$_3$ by the normalized solar spectrum, and then convolved with a gaussian profile with a Full Width at Half Maximum (FWHM) equal to the VLT-ESPRESSO spectral resolution, so that the synthetic spectrum of Titan is showcased at the same spectral resolution as the observed data - again following the procedure described in \cite{Rianço-Silva2024}.

\section{Results and Discussion}
\label{sec:results}

\subsection{Comparing model spectra and observations}
\label{sec:model_vs_data}

\begin{figure*}	\includegraphics[width=\textwidth]{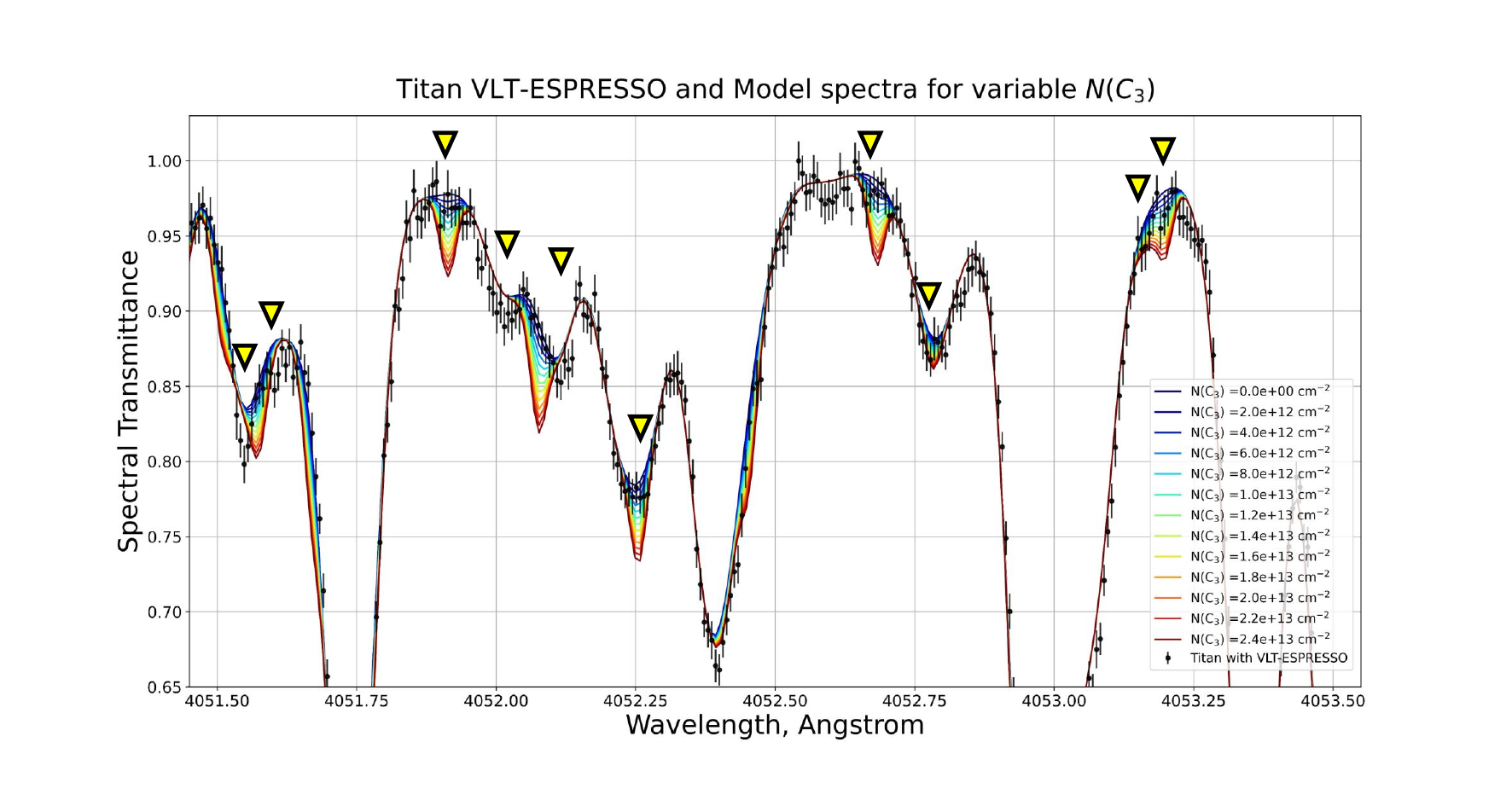}
    \caption{VLT-ESPRESSO normalized spectrum of Titan (black datapoints) at the spectral region of interest for the $\tilde{A}^1 \Pi_u - \tilde{X}^1\Sigma^+_g$ 000-000 band of C$_3$ compared to an array of spectral models of Titan with varying column densities of C$_3$ ($N$). Spectral models based on the \protect\cite{Fan2024} C$_3$ linelist. Telluric transmission obtained from Planetary Spectrum Generator, shifted upwards by 1\% for clarity \protect\cite{Villanueva2018}. The yellow triangles mark the identified spectral features we associate to C$_3$ absorption features, described in table \ref{tab:linelist_c3}.}

    \label{fig:C3_sweeping_original}
\end{figure*}

In order to compare the observed VLT-ESPRESSO spectrum of Titan with the synthetic model spectra with C$_3$ absorption, we have created an array of model spectra by varying the column density of C$_3$, $N$, from 0 to $3 \times 10^{13}$ cm$^{-2}$, with $2 \times 10^{12}$ cm$^{-2}$ steps and assuming $T = 200$K - the Temperature of Titan's atmosphere at an altitude of 400 km \citep{Horst2017}. This comparison is shown in figure \ref{fig:C3_sweeping_original}.

\par Although the most prominent features in this region of the spectrum are the backscattered solar absorption lines, it is noteworthy that several features in the VLT-ESPRESSO spectrum of Titan match those in the simulated Titan spectrum that include modelled C$_3$ absorption lines. A comparison between modelled and observed spectral features is provided in appendix table \ref{tab:linelist_c3}. These features deviate from the normalized solar spectrum, which represents Titan’s baseline spectrum without any C$_3$ absorption. This is particularly evident for the spectral features at 4052.27\AA \hspace{0.1mm}, 4052.77\AA \hspace{0.1mm}  and 4053.19\AA. However, it is also noticeable that in some cases, the C$_3$ absorption lines in the modelled spectra are shifted by few hundredths of \AA ngstrom from the observed spectral features that diverge away from the normalized solar spectrum. This is particularly evident for the spectral features at 4051.60\AA \hspace{0.1mm}, 4051.90\AA \hspace{0.1mm}, 4052.11\AA \hspace{0.1mm} and 4052.67\AA \hspace{0.1mm}. 

\par A possible explanation for this are uncertainties on the C$_3$ linelists used to model the C$_3$ absorption spectrum, taken from \cite{Tanabashi2005}, \cite{Schmidt2014} and \cite{Fan2024}. This is since the C$_3$ linelists were empirically obtained and validated through the observation of interstellar C$_3$ absorption lines using VLT-UVES observations with resolving powers of R $\approx$ 80 000 \citep{Fan2024}. This implies a spectral resolution for past VLT-UVES observations no better than $\Delta \lambda = R/\lambda$, or $\Delta \lambda = $ 0.05\AA \hspace{0.1mm} at $\lambda = $ 4050\AA, which compares to a VLT-ESPRESSO spectral resolution at the UHR mode of  $\Delta \lambda = $ 0.02\AA. It is possible that this observation with improved resolving power requires a correction in some of the wavelengths associated to the centre of some of the C$_3$ lines, as was done in past works following an improvement in spectral resolution compared to the previous literature \citep{Tanabashi2005}, \citep{Schmidt2014}, \citep{Fan2024}.

\subsection{$\Delta\chi^2$ analysis}
\label{section:chi2}
To assess whether non-zero values of modelled C$_3$ absorption actually return an improvement to the quality of the model fit to the VLT-ESPRESSO data, we have calculated a $\Delta\chi^2$ curve to the model fit to the data, as a function of column density of absorbing C$_3$, $N(C_3)$. This analysis follows the approach of \cite{Nixon2020}, where $\chi^2$ is defined by eq. \ref{eq:chi2}, with $S_\lambda$ corresponding to the VLT-ESPRESSO measured Titan spectral flux, $M_\lambda(N)$ corresponding to the modelled Titan flux as a function of C$_3$ column density, $N$, and $\delta S_\lambda$ corresponding to the VLT-ESPRESSO measured uncertainty (i.e., the measured spectral flux at wavelength $\lambda$ by VLT-ESPRESSO is given by $S_\lambda \pm \delta S_\lambda$).

\begin{equation}
    \chi^2(N) = \sum_\lambda \left( \frac{S_\lambda - M_\lambda(N)}{\delta S_\lambda} \right)^2, \hspace{0.25cm} \Delta\chi^2(N) = \chi^2(N) - \chi^2(N = 0) 
    \label{eq:chi2}
\end{equation}

Based on the framework of \cite{Nixon2020}, we calculate the difference of $\chi^2$, $\Delta\chi^2(N)$, by subtracting $\chi^2(N = 0)$ (the $\chi^2$ value for the no C$_3$ case) to each calculated $\chi^2(N)$. If the spectral fit improves with the inclusion of C$_3$ absorption in the model, we would expect $\Delta\chi^2(N)$ to decrease until a strong minimum is reached - which should indicate the $N(C_3)$ for which the best model fit is obtained. It should also be noted that unlike for standard $\chi^2$ analysis, the approach followed by \cite{Nixon2020} is indicative of a good spectral fit for $\chi^2 \sim n$, where $n$ is the number of datapoints used in the fit, set at $n = 1000$.

\par
\begin{figure}	\includegraphics[width=\linewidth]{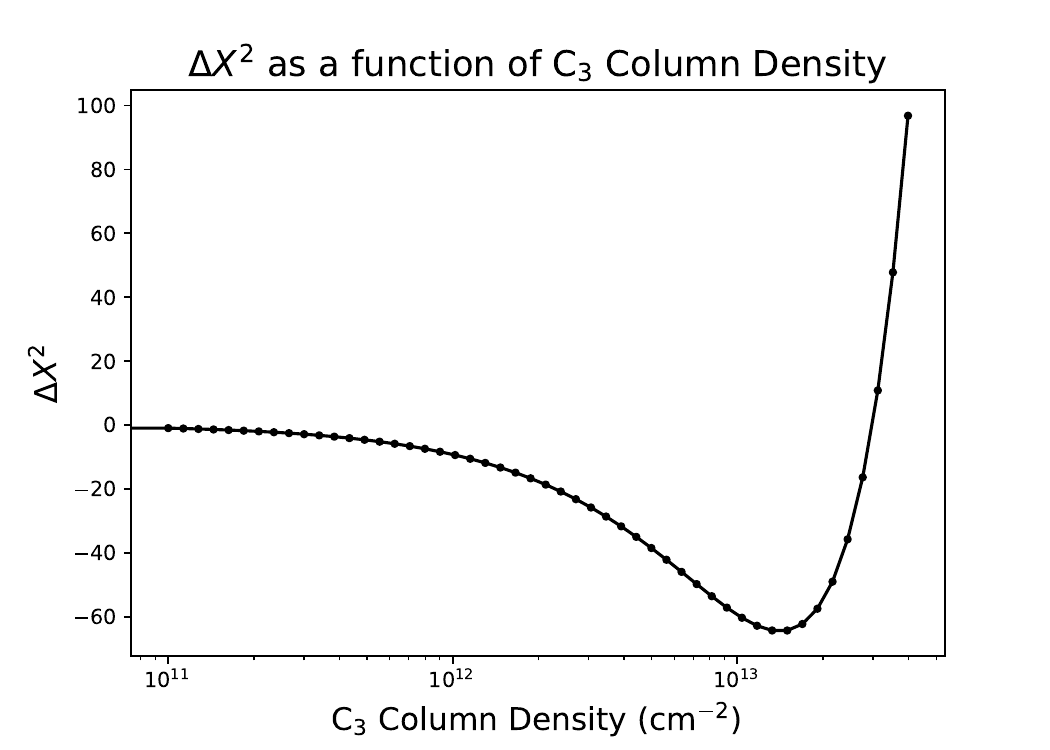}
    \caption{$\Delta\chi^2$ curve as a function of column density of C$_3$, $N(C_3)$, following the approach of \citep{Nixon2020}. This plot shows an increase in the quality of the spectral as the column density approaches $N(C_3) = 1.5 \times 10^{13}$ cm$^{-2}$, suggesting a detection of C$_3$ in Titan.}
    \label{fig:chi2}

\end{figure}

 In figure \ref{fig:chi2} we show the $\Delta \chi^2 (N)$ curve we obtained by calculating the $\Delta\chi^2 (N)$ for a set of 50 values of $N$ spaced in logspace from $N = 10^{11}$ cm$^{-2}$ to $N = 10^{14}$ cm$^{-2}$. The $\chi^2(N = 0)$ in this distribution is 1080, which secures that the model fit to the spectral is good. As would be expected for a C$_3$ detection, we observe a strong minimum emerging in the $\Delta \chi^2 (N)$ curve as $N$ increases, implying that adding C$_3$ absorption to the spectral model improves the quality of the spectral fit (up to an inversion point on this trend, a $\chi^2$ minimum). The $\Delta \chi^2 (N)$ strong minimum is observed at $N(C_3) = 1.5 \times 10^{13}$ cm$^{-2}$, a column density in agreement with the past tentative estimates of C$_3$ abundance in Titan's upper atmosphere \citep{Rianço-Silva2024}.

\par Furthermore, the local minimum's depth in $\Delta\chi^2$ is of -64.2. Again, following the approach of \cite{Nixon2020}, we retrieve this result's significance as $\sqrt{64.2}$, thus yielding a detection significance of C$_3$ in Titan of 8.0$\sigma$.

\subsection{MCMC Likelihood fit to the spectrum}
\label{MCMC_section}

\par In order to further and independently clarify this detection of C$_3$ in Titan and retrieve the atmospheric abundance of C$_3$ in a more robust way, we turn to Bayesian statistics as is standard procedure in atmospheric retrievals based on planetary spectra, in particular for atmospheric retrievals on exoplanet spectra (e.g. \cite{Waldmann2015}, \cite{Taylor2023}, \cite{Xue2024}). Although not widely used for solar system targets, where SNR are often high enough to produce unequivocal detections of individual spectral lines, Bayesian statistical analysis provides a useful tool to examine situations such as this, where absorption features are of the order of magnitude of errorbars \citep{Waldmann2015} as is the case here.

\par Taking as a starting point our simple model spectrum of Titan, $M_\lambda(T, N)$, which considers a single absorbing layer of C$_3$, the 2 free variables that we aim to retrieve from the spectrum are the column density of C$_3$, $N(C_3)$,  and its temperature, $T$. In order to sample and assess the probability distribution of the fits to the multiple models to the spectral data, we use a Markov-Chain Monte Carlo (MCMC) ensamble sampler based on the {\fontfamily{qcr}\selectfont emcee} python package \citep{Mackey2013}.

\begin{figure}	\includegraphics[width=\linewidth]{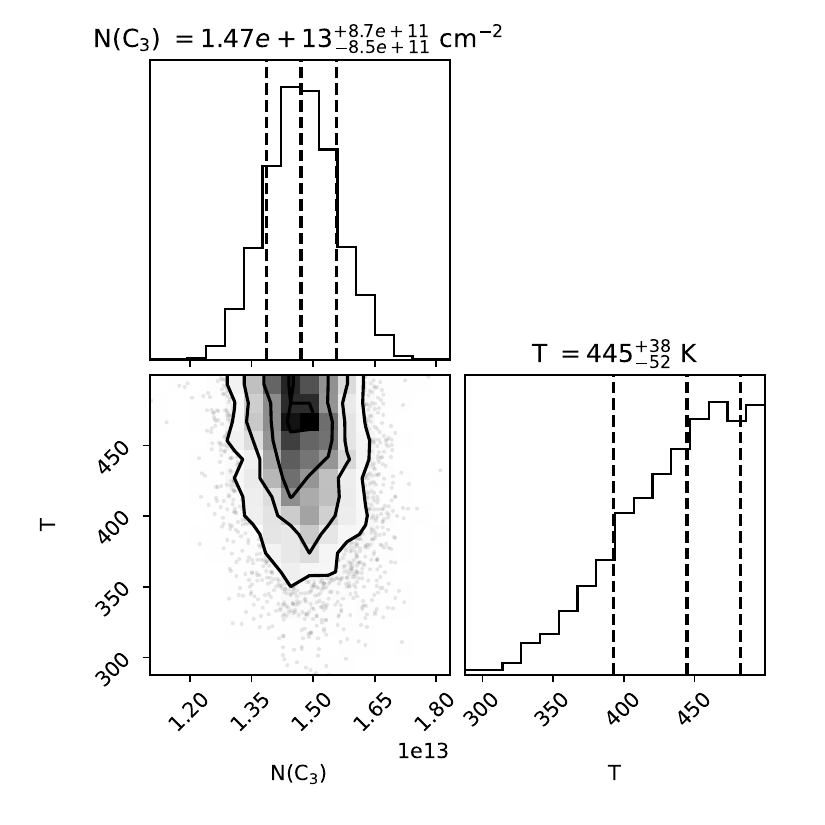}
    \caption{Posterior plots from the MCMC fit to the VLT-ESPRESSO spectrum of Titan. Fit for the column density of C$_3$, ($N$) and for the Temperature ($T$). Retrieved values showcased with 1$\sigma$ errorbars. Contour plots on the 2D histogram correspond to the (0.5, 1, 1.5, 2)-sigma confidence regions for a 2D Gaussian distribution. Histogram y axis correspond to the retrieved probability distribution, unitless.}
    \label{fig:posteriors_original}

\end{figure}

For a MCMC ensamble sampler, with spectral datapoints described by $\lambda, S_\lambda,  \delta S_\lambda$ (respectively, wavelength, measured spectral flux and respective spectral flux error), and fitted variables with $N$ and $T$, the Bayesian argument to the spectral fit is given by eq. \ref{bayesian_argument}

\begin{equation}
    p(N, T | \lambda, S_\lambda, \delta S_\lambda) \propto p(N, T). p(S_\lambda|\lambda, \delta S_\lambda, N, T)
    \label{bayesian_argument}
\end{equation}

\par From eq. \ref{bayesian_argument} we get that the probability distribution function of the retrieved variables $N, T$ applied to the spectral dataset is proportional its prior function $p(N, T)$ and the likelihood function, $p(S_\lambda|\lambda, \delta S_\lambda, N, T)$, following \cite{Mackey2013} and \cite{Waldmann2015}. The priors for $N$ and $T$ were defined as uniform priors with bounds 0 and $10^{14}$ cm$^{-2}$ and 0 to 500 $K$, respectively. These wide prior ranges allow for the MCMC sampler to explore a wide range of realistic values for these variable's parameter space. Also following \cite{Mackey2013} and \cite{Waldmann2015}, the likelihood function can be defined as a log-likelihood function, given by eq. \ref{eq:likelihood}, which we aim to maximize.

\begin{equation}
   \log(\mathcal{L}) = - \frac{1}{2}\sum_\lambda\left[ \frac{(S_\lambda - M_\lambda(N, T))^2}{(\delta S_\lambda)^2}\right] - \log\left(\sqrt{2\pi(\delta S_\lambda)^2}\right)
   \label{eq:likelihood}
\end{equation}

\label{sec:mcmc}

\par This MCMC likelihood fit to the VLT-ESPRESSO spectrum of Titan was run for 5000 iterations, having returned the posterior plots of figure \ref{fig:posteriors_original} . Unlike the $\chi^2$ curve shown in section \ref{section:chi2}, retrieval posterior plots do not show the quality of the spectral fit in units of noise standard deviations. Instead, these correspond to probability distribution histograms, showcasing the variable ($N$ and $T$) values for which the spectral fit log-likelyhood function is maximised. It is their distribution as a function of the assessed free parameters ($N$ and $T$) that yields the most likely value for these variables (the histogram's median) as well as the retrieved value's errorbars (based on the standard deviations of the probability distribution), \citep{Waldmann2015}. These posterior plots showcase a good consistency in the model's fit to the VLT-ESPRESSO spectrum of Titan for C$_3$ column density, allowing constraints to be placed on the abundance of C$_3$. Based on the posterior probability distribution to the MCMC fit and using the original \cite{Fan2024} C$_3$ linelist, we retrieve a column density of C$_3$ in Titan of $N = \left(1.47 \pm 0.09 \right) \times 10^{13}$ cm$^{-2}$ and an associated temperature of $T = 445\hspace{0.5mm} ^{+\hspace{0.5mm}38}_{-\hspace{0.5mm}52}$ K, for 1$\sigma$ uncertainties, as shown in the histograms of figure \ref{fig:posteriors_original}. 

\par The retrieved values for 3$\sigma$ uncertainties are $N = \left(1.47 \hspace{0.5mm}^{+\hspace{0.5mm}0.25}_{-\hspace{0.5mm}0.26}\right) \times 10^{13}$ cm$^{-2}$ and an associated temperature of $T = 445 \hspace{0.5mm} ^{+\hspace{0.5mm}57}_{-\hspace{0.5mm}134}$ K. For 5$\sigma$ uncertainties, we get $N = \left(1.47 \pm 0.30\right) \times 10^{13}$ cm$^{-2}$ and an associated temperature of $T = 445\hspace{0.5mm} ^{+\hspace{0.5mm}58}_{-\hspace{0.5mm}171}$ K. The posterior plot histogram showcases nearly symmetrical errorbars for the retrieval of C$_3$ column density, with errorbars confidently excluding the $N = 0$  scenario (equivalent to no C$_3$ on Titan). This further suggests the detection of C$_3$ on Titan's upper atmosphere with a confidence above 5$\sigma$, which is the standard to claim a new detection.

\par The histogram concerning the retrieval of $T$ is not symmetrical, presenting a heavier tail skewed to higher temperature values revealing a challenging constraint in the temperature on this retrieval. Alongside the larger relative uncertainties in the temperature (uncertainties close to 50\% for the retrieved $T$ at 5$\sigma$), this skewed distribution suggests a lower sensitivity to temperature on this fit. It may result as a consequence of the approximations applied to this model, in particular regarding thermal equilibrium and a constant temperature across the entire absorbing layer of C$_3$ which deviates from observed temperature profiles of Titan \citep{Willacy2022}. It may also stem from the aforementioned putative errors in some C$_3$ line positions in \cite{Fan2024} which would become more evident at these unprecedented spectral resolutions. Indeed, given the wide parameter range which this retrieval allowed for $T$ (0K to 500K), it can be argued that the retrieval could have tended towards higher temperatures in order to enable substantial line broadening which would approximate the model fit to the slightly offset spectral features. In appendix \ref{Appendix_Adapted_Linelist} we test whether a slightly adapted linelist for C$_3$, based on the original \cite{Fan2024} linelist, provides a better constrained posterior plot for $T$. This is verified in the posterior plots of figure \ref{fig:posteriors_original_vs_adapted} - pointing towards a possible improvement of the available C$_3$ linelist, based on the fact that C$_3$ is clearly detected from an independent method, as was the $\chi^2$ curve using the original linelist. These small shifts that we suggest applying to the original \cite{Fan2024} C$_3$ linelist are provided at the last column of table \ref{tab:linelist_c3}. Despite this, the retrieved temperature's large uncertainties broadly matches the maximum expected temperature for the altitude where the highest abundance of C$_3$ was predicted to occur - at an altitude of 400 km \citep{Herbad2013} \citep{Dobrijevic2016}, with a temperature of about 200K \citep{Horst2017}.

\par
\begin{figure}	\includegraphics[width=6cm]{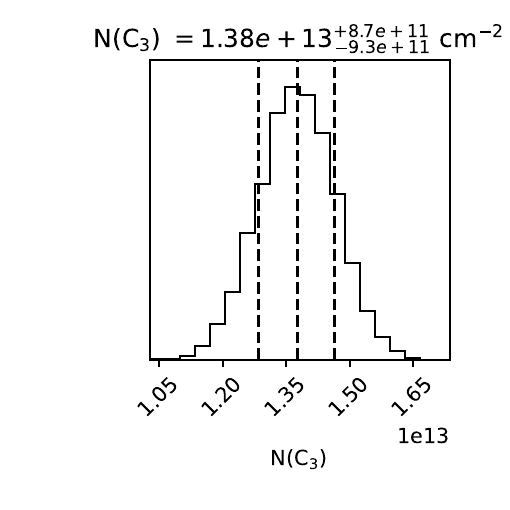}
    \caption{Posterior plots from the MCMC fit to the VLT-ESPRESSO spectrum of Titan. Fit for the column density of C$_3$, ($N$) for a fixed Temperature value of $T = 200$K. Retrieved value showcased with 1$\sigma$ errorbar. Histogram y axis correspond to the retrieved probability distribution, unitless.}
    \label{fig:1d_retrieval}

\end{figure}

 Nonetheless, these Bayesian retrievals also enable us to fit $T$ to more physically motivated values (in our case, T = 200K, from \cite{Horst2017}), which is helpful to test if even without the aforementioned non-physical line broadening enabled by the retrieval, the fit is still able to provide a 5$\sigma$ detection of C$_3$. The result of such a fit is showcased in figure \ref{fig:1d_retrieval}. It also shows a well constrained retrieval for $N(C_3)$, with $N(C_3) = \left(1.38\pm 0.09\right) \times 10^{13}$ cm$^{-2}$ at 1$\sigma$, $N(C_3) = \left(1.38^{+\hspace{0.5mm}0.26}_{-\hspace{0.5mm}0.27}\right) \times 10^{13}$ cm$^{-2}$ at 3$\sigma$ and $N(C_3) = \left(1.38^{+\hspace{0.5mm}0.36}_{-\hspace{0.5mm}0.35} \right) \times 10^{13}$ cm$^{-2}$ at 5$\sigma$, for $T$ fixed at 200K. This retrieved value matches the results obtained by the $\chi^2$ fit and the MCMC retrieval for $N$ and $T$, which further confirms the detection described above, when fixing the temperature value, as the $N(C_3) = 0$ scenario is excluded at 5$\sigma$.

\par
\begin{figure*}	\includegraphics[width=\textwidth]{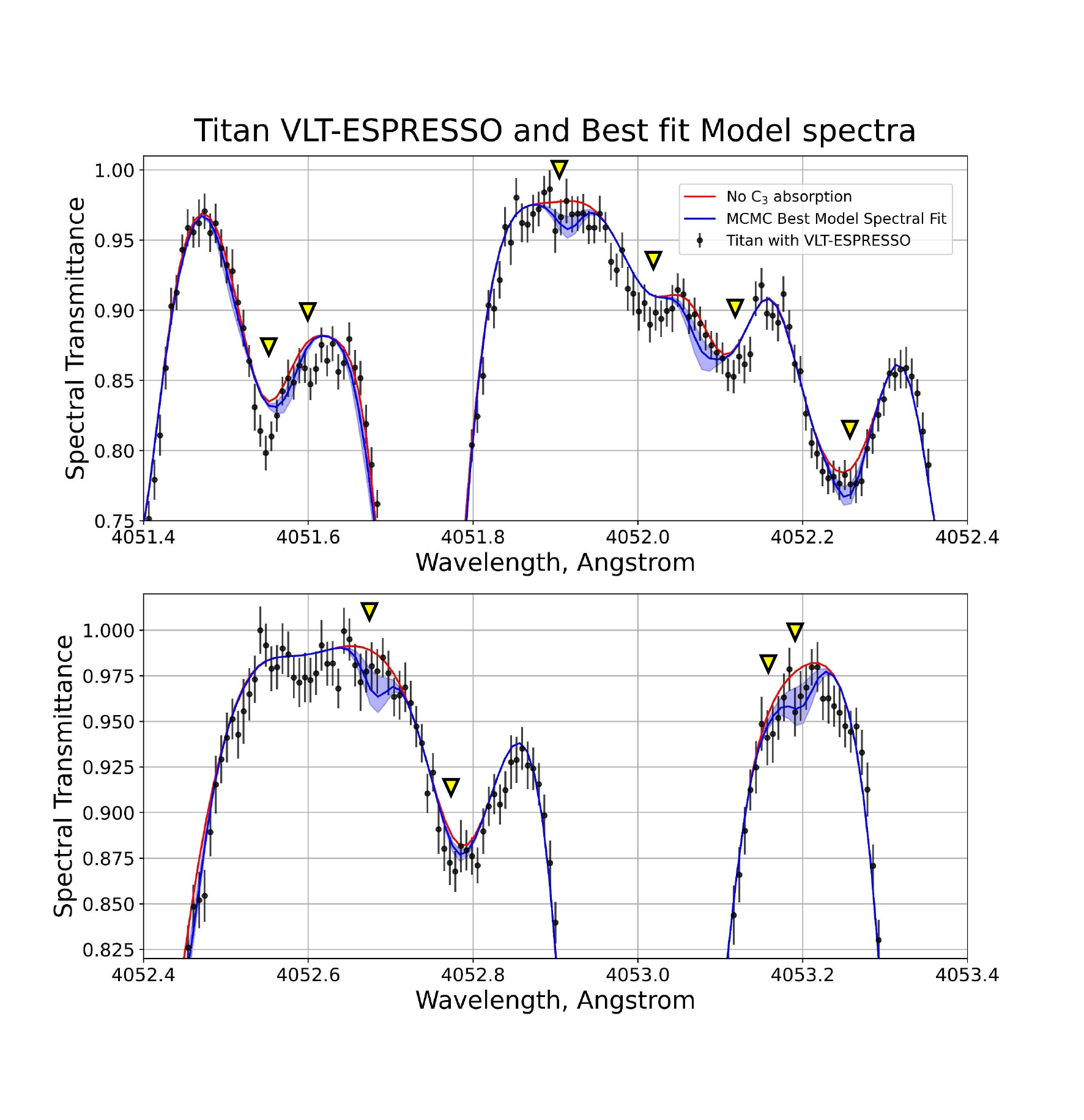}
    \caption{VLT-ESPRESSO normalized spectrum of Titan (black datapoints) at the spectral region of interest for the $\tilde{A}^1 \Pi_u - \tilde{X}^1\Sigma^+_g$ 000-000 band of C$_3$ compared to the backscattered solar spectrum (a proxy for Titan's spectrum with no C$_3$ absorption, in red) and to the best-fit model obtained from the MCMC likelihood fit, associated with the retrieved values for $N(C_3)$ and $T$, in blue. The shaded blue area corresponds to the best fit model uncertainty based on the retrieved $N(C_3)$ and $T$ values with 3$\sigma$ errorbars. The yellow triangles mark the identified spectral features we associate to C$_3$ absorption features, described in table \ref{tab:linelist_c3}.}    
    \label{fig:best_fit}
\end{figure*}

 A plot showcasing the best fit model for this VLT-ESPRESSO spectrum of Titan obtained from the best fit parameter values for $N(C_3)$ and $T$, 3$\sigma$ respective uncertainties, compared with the backscattered solar spectrum (a proxy for Titan's spectrum with no C$_3$ absorption), is shown in figure \ref{fig:best_fit}. This best fit plot with respective uncertainties further highlights the observed spectrum's deviation from the solar spectrum on the wavelengths subject to C$_3$ absorption - further confirming the detection of C$_3$ on Titan. In the appendix \ref{Appendix_Residuals}, particularly in figure \ref{fig:Original_LL_residuals}, we showcase the residual plots relative to the observed spectrum and the MCMC best fit spectral model - and one residuals plot relative to the spectral model with no C$_3$. As expected, the best fit MCMC spectral model (with C$_3$) improves the fit, decreasing the residual's deviation to closer to zero.

\par Interestingly, the retrieved column density for C$_3$ in Titan matches the estimated column density value found at the first tentative detection of C$_3$ in Titan with VLT-UVES in \cite{Rianço-Silva2024}, and within the predicted upper and lower limits for $N(C_3)$ from \cite{Dobrijevic2016}, which suggested C$_3$ column densities across Titan's mesosphere from $5 \times 10^{12}$ cm$^{-2}$ to $5 \times 10^{13}$ cm$^{-2}$.

\section{Conclusions}
\label{sec:Conclusions}

In this work, we have obtained and analysed VLT-ESPRESSO high resolution visible spectra of Titan - which are to date, the visible spectra of Titan at the highest spectral resolution ever obtained. This unprecedented spectral resolution has enabled an increased sensitivity within the short wavelength range of interest for the search for C$_3$ absorption features, mostly in the 4050\AA \hspace{0.1mm} to 4055\AA \hspace{0.1mm} range, as compared to previous high resolution observations of Titan \citep{Rianço-Silva2024}. Most importantly, this increased spectral resolution and sensitivity enabled by VLT-ESPRESSO disentangles more effectively the underlying and deeper solar spectral features from possible C$_3$ absorption features on Titan's spectrum, which was a considerable issue at lower resolutions \citep{Rianço-Silva2024}.

\par Hence, and despite the significantly deeper solar spectral features, comparison with a set of model spectra of Titan's atmosphere with increasing abundances of C$_3$ has enabled identifying matching features that deviated away from the \cite{Kurucz2006} solar spectrum. We identify 10 absorption features deviating by more than 1$\sigma$ away from the solar spectrum, and matching the depth and position of C$_3$ absorption features. Although some of these observed non-solar spectral features may appear to slightly deviate from the predicted C$_3$ absorption line position, from \cite{Fan2024}, in all cases the deviation is within this linelist errorbar of 0.05\AA  - see table \ref{tab:linelist_c3}.

\par Despite the compelling case for the detection of C$_3$ in Titan with the 10 matching absorption features described above, none of the individual potential C$_3$ features deviates from the $N = 0$ spectrum by $5\sigma$, which would be required to claim a detection just by observing one individual line. This was addressed by performing 2 independent statistical analysis to the model fit to the VLT-ESPRESSO spectrum. The first consisted in calculating the $\chi^2$ curve of the model fit as a function of C$_3$ column density (following the approach used in \cite{Nixon2020}). This yielded a strong minimum at $N = 1.5 \times 10^{13}$ cm$^{-2}$ consistent with a C$_3$ detection at 8$\sigma$. The second approach consisted in applying a simple Bayesian MCMC retrieval \citep{Waldmann2015} \citep{Mackey2013}, fitting the spectral model to the observed spectrum. With this approach, it was possible to retrieve the total C$_3$ column density, $N$, and temperature, $T$, that would provide the best model fit to the observed spectrum. Crucially, this fit has enabled retrieval of a C$_3$ column density of $N = \left(1.38 \hspace{0.5mm}^{+\hspace{0.5mm}0.36}_{-\hspace{0.5mm}0.35}\right) \times 10^{13}$ cm$^{-2}$ (at 5$\sigma$), hence ruling out the $N = 0$ scenario (no C$_3$ on Titan) at 5$\sigma$, enabling a detection of C$_3$ on Titan. In spite of the larger relative errorbars and poorly constrained posterior (which does not show a strong peak), the retrieved temperature value for the C$_3$ absorbing layer, $T = 422\hspace{0.5mm} ^{+\hspace{0.5mm}58}_{-\hspace{0.5mm}171}$ K, is broadly in agreement with the expected Temperature in Titan's mesosphere ($T \sim$ 200K), where \cite{Dobrijevic2016} predicts that C$_3$ is more abundant. 

\par
\begin{figure}	\includegraphics[width=0.95\linewidth]{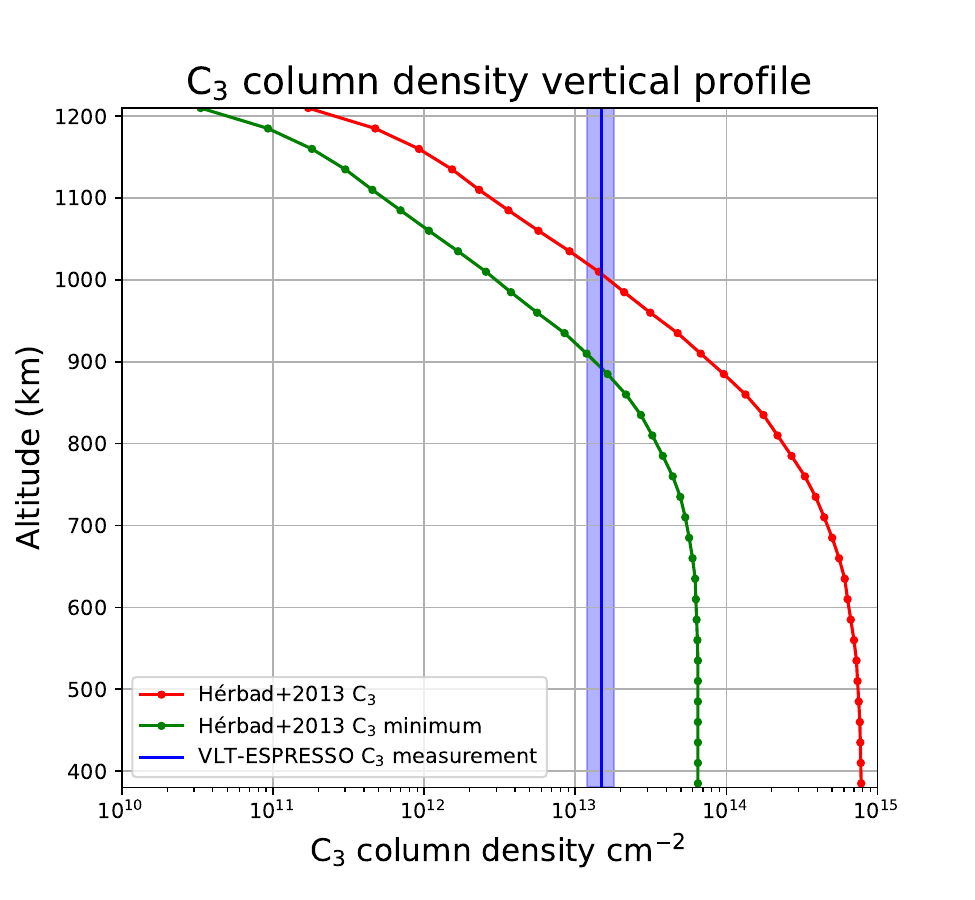}
    \caption{Vertical profile of C$_3$ column density (from the top of the atmosphere down to the altitude shown in the y axis), obtained from the integration in altitude of the \protect\cite{Herbad2013} predicted vertical profile of C$_3$, for the median profile obtained from the uncertainty propagation study of C$_3$ (in red) and for the 1st 20-quantile distribution (in green, described as C$_3$ minimum predicted profile). The blue line and errorbar corresponds to the C$_3$ column density retrieved on this work, with the 5$\sigma$ associated errorbar.}
    \label{fig:Vertical_Profile_C3}

\end{figure}

 The retrieved value of C$_3$ column density on Titan's upper atmosphere in this work can be compared to the predicted C$_3$ abundance profiles from previous modelling studies of Titan's atmospheric chemistry \citealt{Herbad2013, Dobrijevic2016}. Taking the predicted vertical profile of C$_3$ mixing ratio from figure 4 of \citep{Herbad2013} and integrating it over the multiple altitude layers, using the \citep{Teanby2009} atmospheric density profile, we have obtained the equivalent prediction of column density of C$_3$ as a function of altitude, down to 400 km of altitude, which is shown in figure \ref{fig:Vertical_Profile_C3}. It is worth mentioning that the \citep{Herbad2013} vertical abundance profiles are sourced from 900 MCMC runs of the photochemical model, for which a median profile and 20-quantile profiles are obtained. The median predicted C$_3$ column density down to 400 km of altitude from the \citep{Herbad2013} photochemical model is of $7.8 \times 10^{14}$ cm$^{-2}$ (column density vertical profile shown in red), 50 times higher than our C$_3$ column density measurement with VLT-ESPRESSO. However, for the 1st 20-quantile of the predicted C$_3$ distribution (column density vertical profile shown in green), the predicted column density of C$_3$ down to 400 km is of $6.5 \times 10^{13}$ cm$^{-2}$, a value 3 times higher than our VLT-ESPRESSO measurement. The observed discrepancy between our observed value and the lower limit for the \citep{Herbad2013} photochemical model prediction of a factor or 3 is significantly smaller than the prediction's uncertainty window (the model's median profile compared to the 1st 20-quantile) which spans over a factor of 12. However, this slight discrepancy but could point towards chemical processes unaccounted for in \citep{Herbad2013} that could result in a slight C$_3$ depletion in Titan's upper atmosphere. Updates to Titan's hydrocarbon chemistry have been further explored recently \citep{Loison2019}, \citep{Willacy2022}, but no updated C$_3$ vertical profiles have been published since \citep{Herbad2013}, hence we suggest future modelling works to explore updates to C$_3$ photochemistry in Titan.  

\par In order to further constrain the vertical profile of C$_3$ in Titan's mesosphere and draw direct comparisons between its observed vertical profile with other photochemical species and photochemical model predictions, we suggest using a more thorough radiative transfer model of Titan's atmosphere, with vertical resolution, rather than a simple 1-layer model as used in this work. This 1-layer model can be thought of as an oversimplification of the line formation, treating it as a "screen" on the spectrum absorbing only at a fixed altitude. However, in order to accomplish this altitude-dependent spectrum, more information regarding the C$_3$ electronic transitions is required when compared to the available information for the 4050 \AA \hspace{0.1mm} absorption band \citep{Fan2024} (e.g: The air-broadened line width, and pressure shifts in line position). Nonetheless, the fact that C$_3$ absorption occurs so high in Titan’s atmosphere (above 300km) at a low-pressure environment causes Doppler-broadening to dominate the line shape, making a Gaussian line profile a reasonable assumption \citep{Tennyson2014}). One such complete linelist of C$_3$ was recently published by ExoMol \citep{Lynas-Gray2024} for C$_3$ mid-infrared transitions for wavelengths longward of 5$\mu$m. Hence we recommend further search and modelling efforts of C$_3$ absorption on Titan on observational data spanning mid-infrared wavelengths, such as those recently published based on JWST-MIRI \citep{Nixon2025}. 

\par To conclude, we further highlight the great usefulness of instruments originally designed for exoplanet research, such as VLT-ESPRESSO \citep{Pepe2021}, or even techniques developed to address the low SNR nature of exoplanet spectra, such as the Bayesian MCMC likelihood spectral retrievals \citep{Waldmann2015}, when applied to Solar System targets and science cases. On the advent of a golden era of technical and scientific development aimed at exoplanet research (e.g., with future telescopes and instruments such as Ariel \citep{ariel} and ELT-ANDES \citep{elt-andes}) it is also worth taking a look back at our own cosmic backyard and revisit many of its mysteries yet to be solved with the light of this new knowledge.

\section*{Acknowledgements}

This work was supported by Fundação para a Ciência e Tecnologia (FCT) of reference PTDC/FIS-AST/29942/2017, through national funds and by FEDER through COMPETE 2020 of reference POCI-01-0145-FEDER-007672, and through the research grants UIDB/04434/2020, UIDP/04434/2020 and UID/04434/2025. RRS acknowledges funding through the FCT fellowship grant 2024.02527.BD. This study was based on observations collected at the European
Organisation for Astronomical Research in the Southern Hemisphere
under ESO programme 114.277N.

\section*{Data Availability}

The observational data used in this study is publicly available at the ESO Science Archive, associated to the observing program 114.277N. We further add, as supplementary materials to this paper, the spectral data used in this study (Titan's spectrum and the model spectra used in this analysis).



\bibliographystyle{mnras}
\bibliography{example} 




\appendix

\section{Spectral Data Reduction}\label{Appendix - DR}
\begin{figure}	\includegraphics[width=\linewidth]{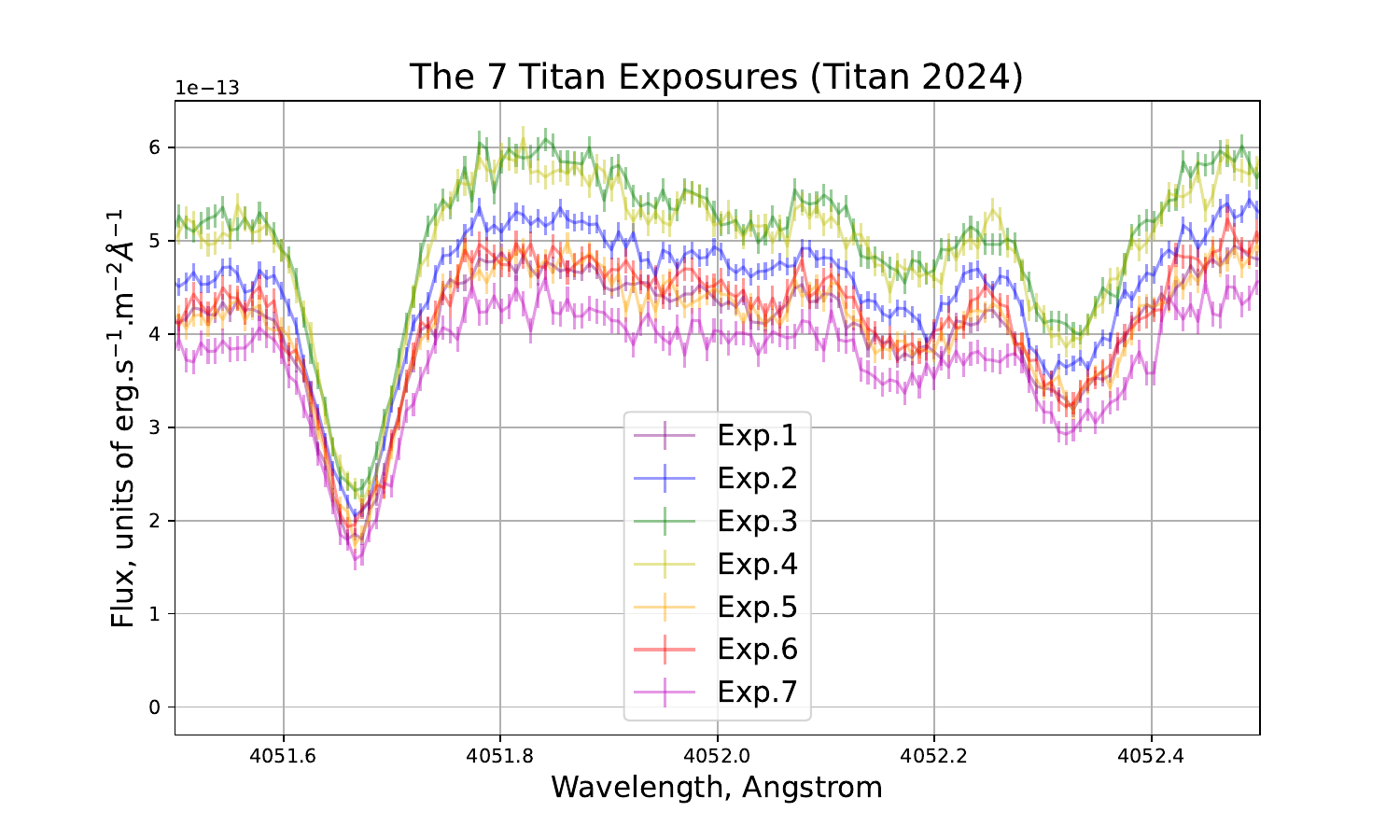}
    \caption{ 7 spectral exposures of Titan obtained by VLT-ESPRESSO, each 20 minutes long (zoomed into a narrow spectral region, for clarity).}
    \label{fig:7_spectral_exposures}

\end{figure}

Here we show in detail the steps taken to perform the data reduction of the spectral data obtained from the observations with VLT-ESPRESSO. As mentioned in section \ref{sec:Observations}, we have obtained 7 spectral exposures of Titan, each 20 minutes long, encompassing a total observation time of 2h20min. Over the course of this time, Titan is observed between airmasses of 1.15 and 1.80, leading to distinct telluric extintions on the observed spectrum over the course of the night. Figure \ref{fig:7_spectral_exposures} shows a comparison between the differing spectra originated on each of the 7 exposures, highlighting the distinct continua across exposures taken when Titan was observed at distinct airmasses.
\par
\begin{figure}	\includegraphics[width=\linewidth]{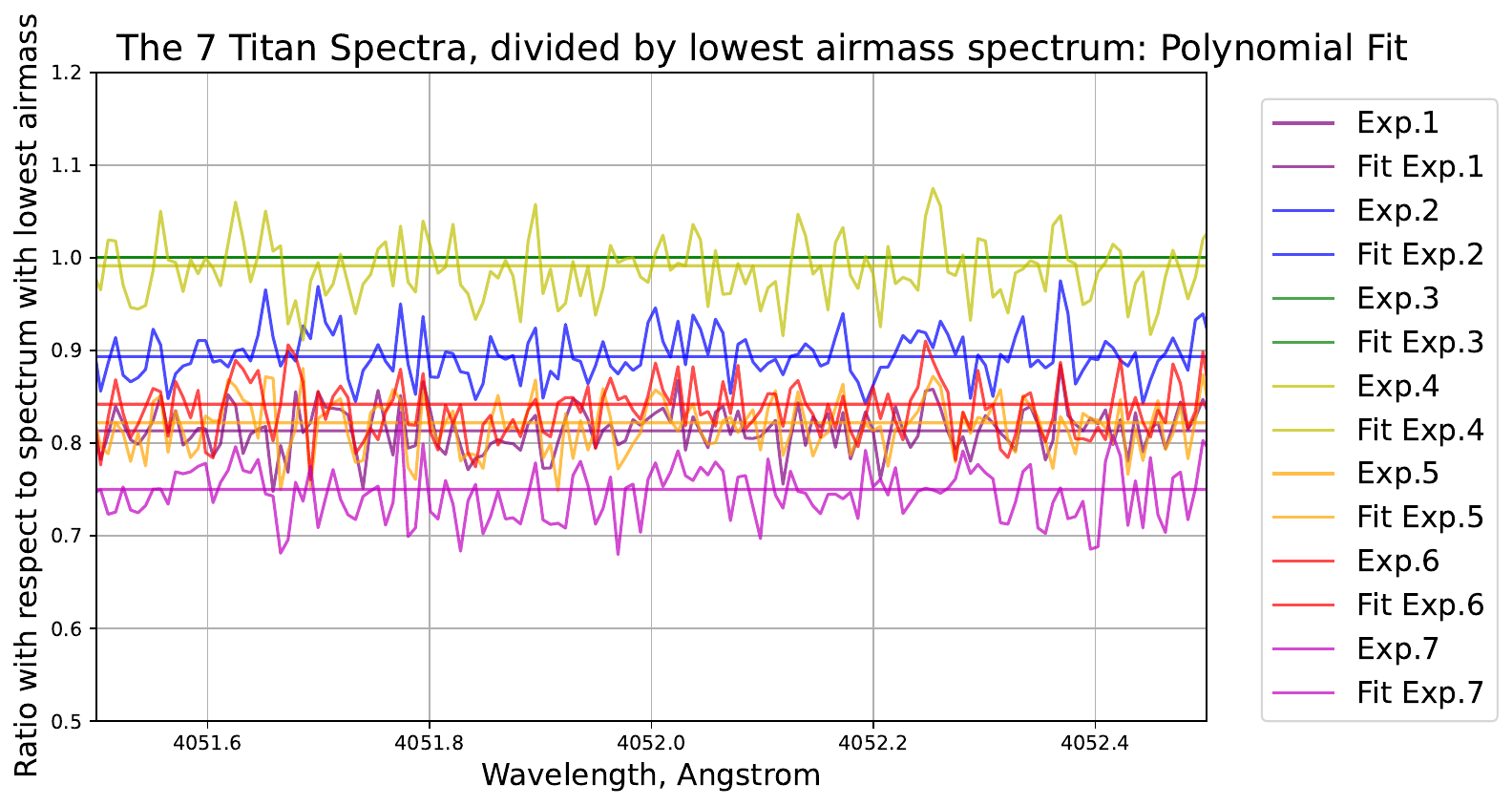}
    \caption{7 spectral exposures of Titan obtained by VLT-ESPRESSO divided by the spectral exposure observed with the lowest airmass (3rd exposure), and their respective 3rd-degree polynomial fits (zoomed in the same spectral region as fig. \ref{fig:7_spectral_exposures} for clarity).}
    \label{fig:7_spectral_exposures_polyfit}

\end{figure}

\begin{figure}	\includegraphics[width=\linewidth]{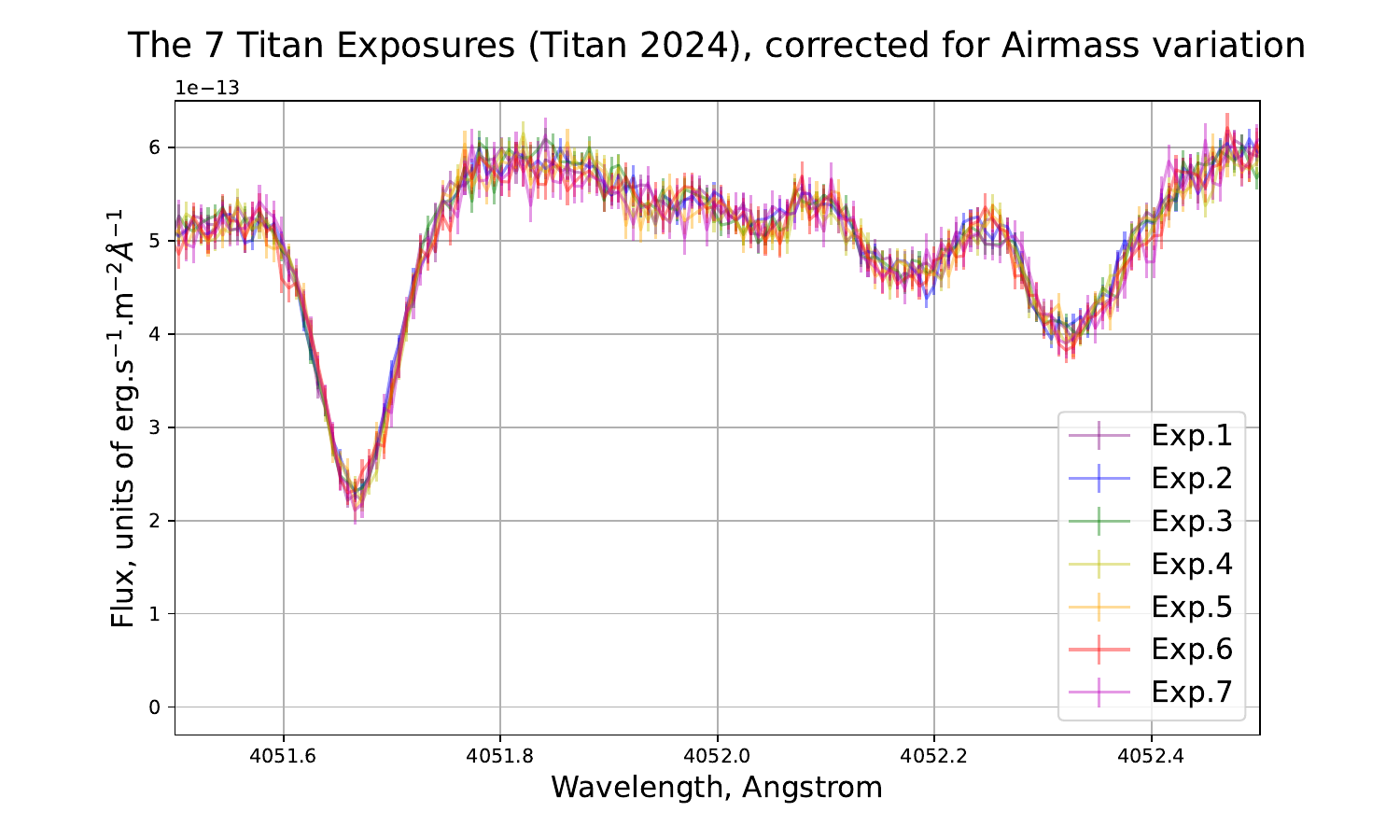}
    \caption{7 spectral exposures of Titan obtained by VLT-ESPRESSO corrected for different airmass extinctions, (zoomed in the same spectral region as fig. \ref{fig:7_spectral_exposures} for clarity).}
    \label{fig:7_spectral_exposures_Airmass_corrected}

\end{figure}

 To correct this, each spectral exposure was divided by the spectral exposure with the lowest airmass (the 3rd exposure), and this ratio was fit by a 3rd-degree polynomial across the entire spectrum, as shown in figure \ref{fig:7_spectral_exposures_polyfit}, which zooms in on the same short wavelength range as figure \ref{fig:7_spectral_exposures}. Each spectral exposure was then divided by the polynomial fit to their respective ratio with respect to the exposure with the lowest airmass, allowing to correct the continuum differences across spectral exposures - allowing the distinct spectral exposures to share the same continuum, as shown in figure \ref{fig:7_spectral_exposures_Airmass_corrected}.
\par
\begin{figure}	\includegraphics[width=\linewidth]{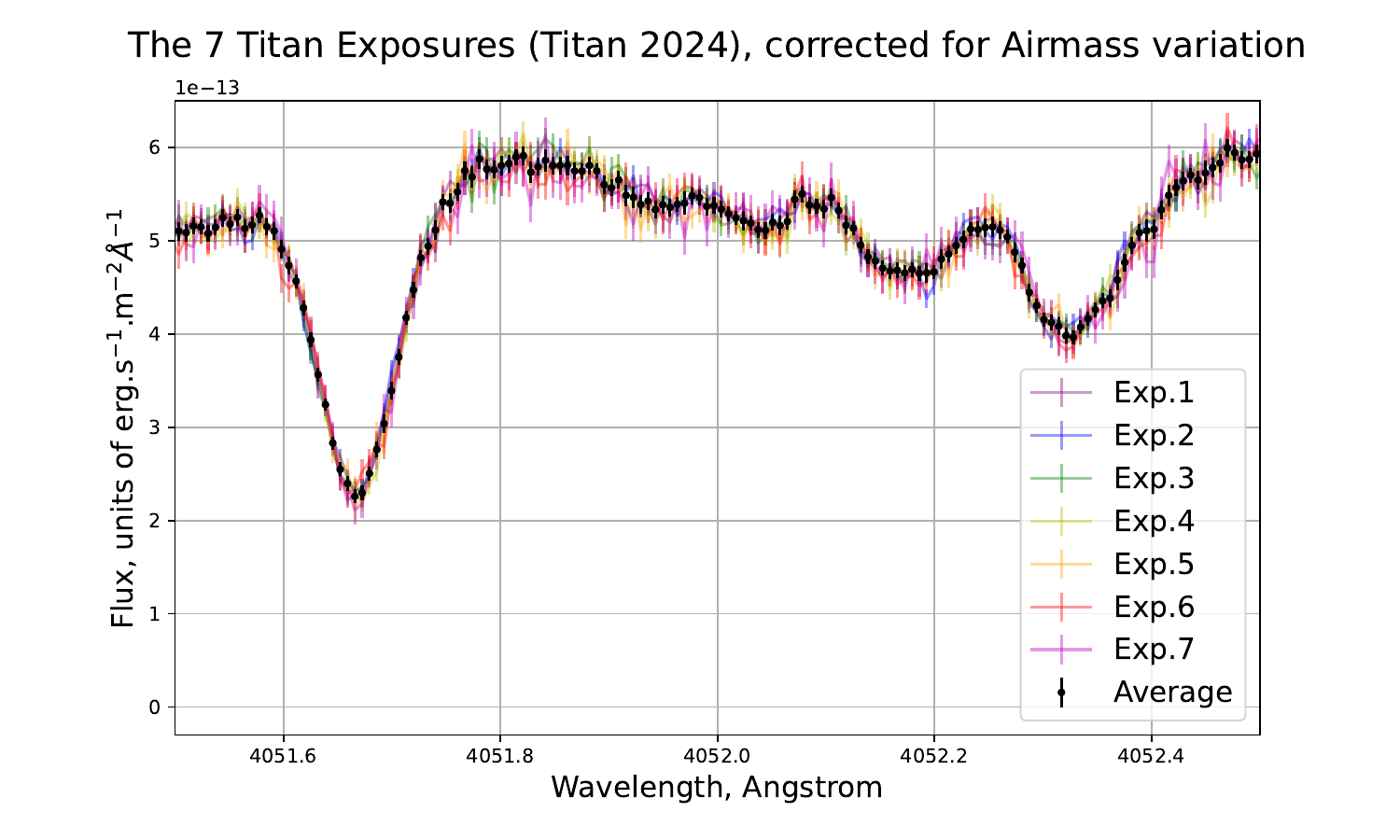}
    \caption{7 spectral exposures of Titan obtained by VLT-ESPRESSO corrected for different airmass extinctions with spectral average and respective errorbars (zoomed in the same spectral region as fig. \ref{fig:7_spectral_exposures} for clarity).}
    \label{fig:7_spectral_exposures_Airmass_corrected_average}

\end{figure}

\begin{figure}	\includegraphics[width=\linewidth]{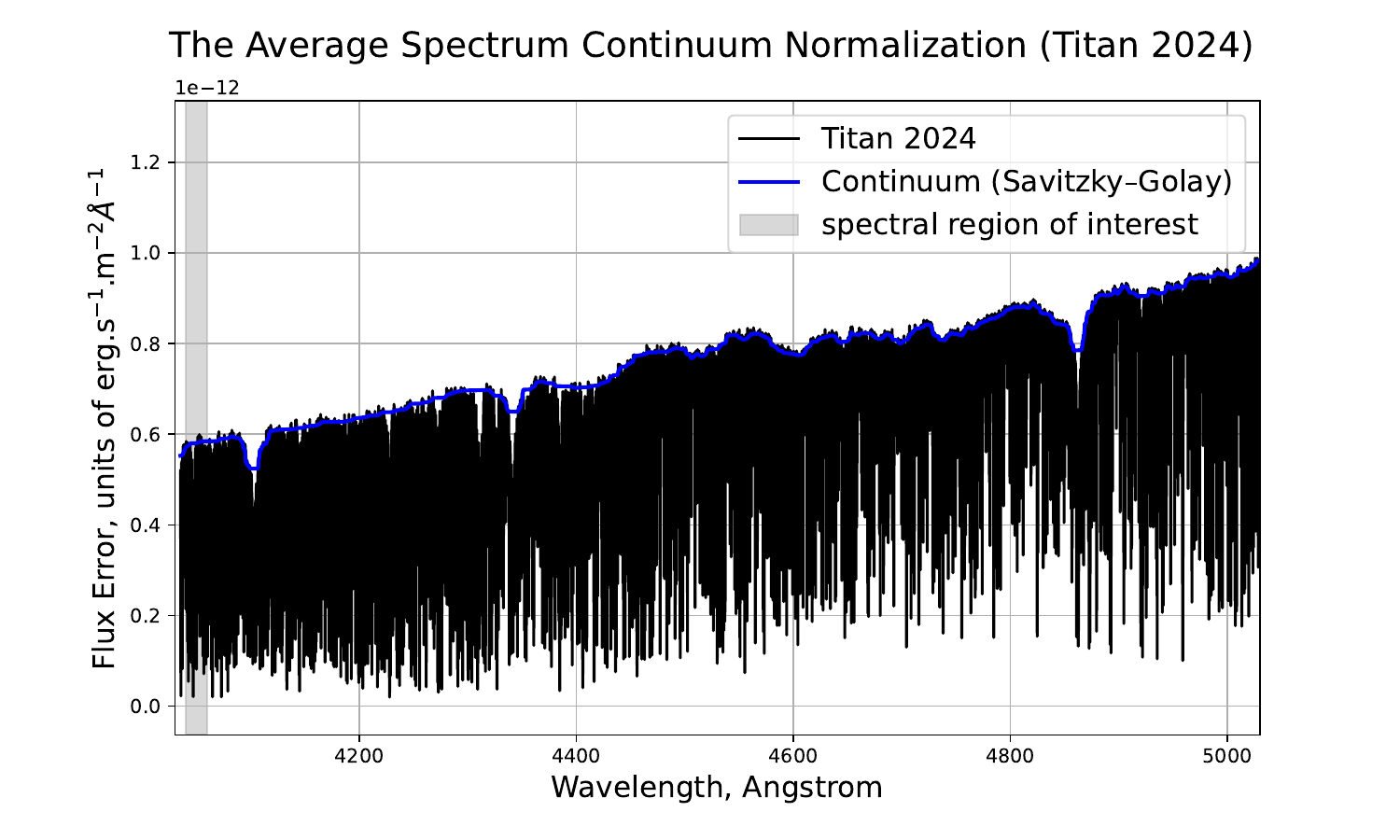}
    \caption{Continuum extraction from the average spectrum of Titan with a Savitzky-Golay filter to enable the normalization of the spectrum - across a larger spectral region, for clarity. The spectral region of interest for this study (4040\AA \hspace{0.1mm} to 4060\AA) is shown in gray.}
    \label{fig:spectrum_normmalization_SG}

\end{figure}

\begin{figure}	\includegraphics[width=\linewidth]{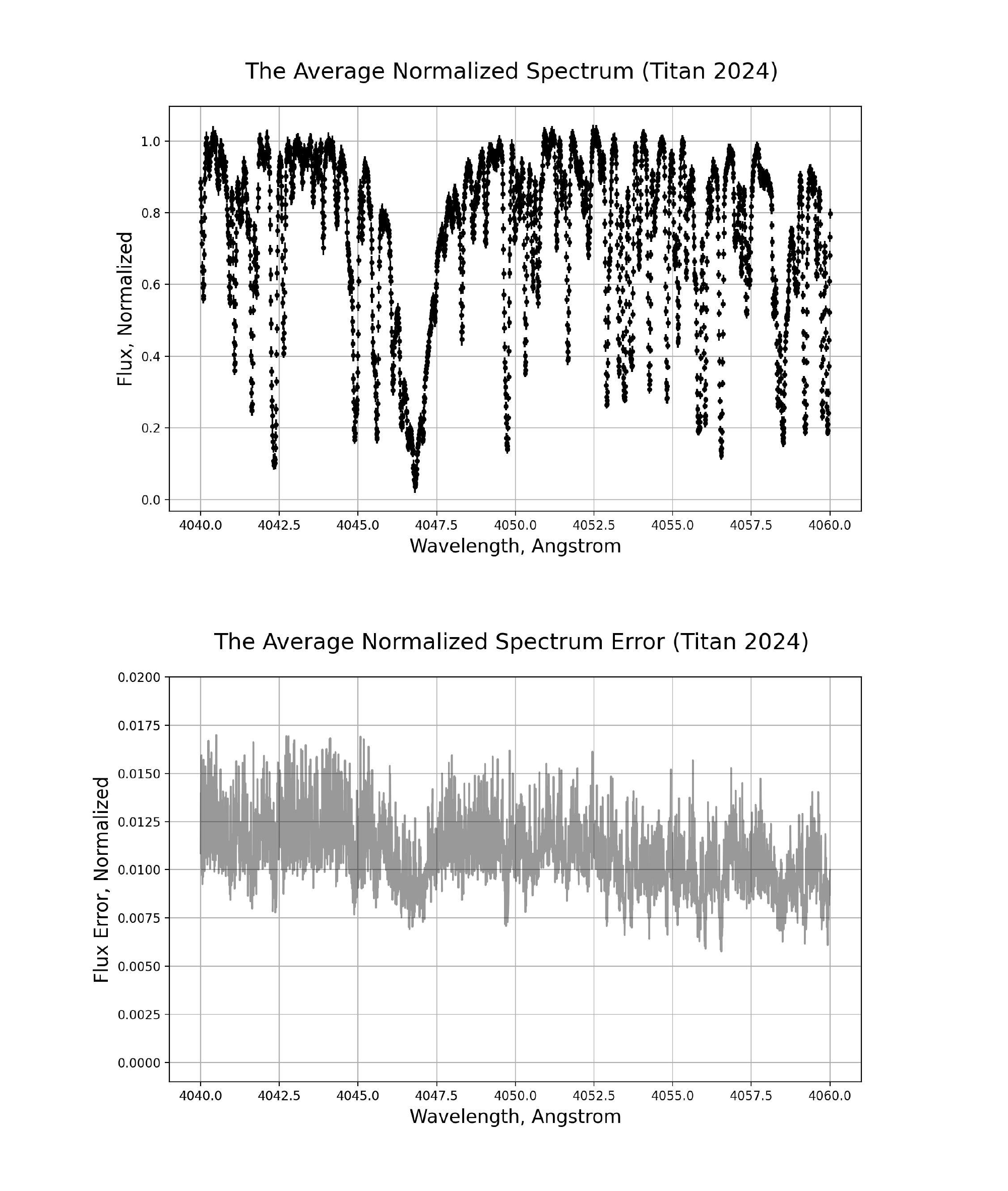}
    \caption{Top: Average Normalized Spectrum of Titan on the spectral region of interest for the search for C$_3$ (4040\AA \hspace{0.1mm} to 4060 \AA). Bottom: Spectral flux errors of the averaged, normalized spectrum as a function of wavelength in the spectral region of interest for the search of C$_3$.}
    \label{fig:plot_67_errors}

\end{figure}

 After sharing a similar spectral continuum, the spectral exposures were averaged into a single spectrum of Titan, with increased SNR, as shown in black, in figure \ref{fig:7_spectral_exposures_Airmass_corrected_average}. In figure \ref{fig:spectrum_normmalization_SG} we show the obtention of the spectral continuum using the Savitzky-Golay filter \citep{Savitzky-Golay} - across a larger spectral region, for clarity - used to normalize the average spectrum of Titan. Finally, in the joint plot of figure \ref{fig:plot_67_errors}, we show the final averaged and normalized spectrum of Titan on top, in the region of interest for the search for C$_3$ (4040\AA \hspace{0.1mm} to 4060 \AA). In the bottom of figure \ref{fig:plot_67_errors} we plot the spectral flux errors of the averaged, normalized spectrum shown on the top of this figure. We observe that despite some modulation that reflects that of the spectral flux (indicating that uncertainties are dominated by photonic error), the spectral error does not vary widely as a function of wavelength on the spectral section used on this study.

\section{Derivation of Airmass Integration on Titan's visible disk}
\label{Appendix}
Here we derive the expression and value of the 2-way airmass, integrated across the entirety of Titan's observed disk.
\par  We consider Titan as a fully illuminated sphere, observed as a resolved disk. Let's consider spherical coordinates for Titan. We define the polar angle, $\theta$, as the angle between a given position on Titan's observed sphere and the sub-Earth point on Titan. A surface element of the observed sphere of Titan is given by $dS$ in eq. \ref{dS}, whereas due to its spherical geometry, the apparent surface element of the sphere projected onto the observed disk is given by $dS_{APP} = \cos(\theta) dS$, or by eq.\ref{dSapp}.

\begin{equation}
    dS = R^2 \sin(\theta)d\theta d\phi
    \label{dS}
\end{equation}

\begin{equation}
    dS_{APP} = R^2 \sin(\theta)\cos(\theta)d\theta d\phi
    \label{dSapp}
\end{equation}

\par Considering Titan as a lambertian reflector - rather than a perfect reflector - we shall include a further $\cos(\theta)$ term on $dS_{APP}$, to account lambertian scatter, as

\begin{equation}
    dS_{APP} = R^2 \sin(\theta)\cos^2(\theta)d\theta d\phi
\end{equation}

We define "Airmass", $A(\theta)$, as the factor by which the Titan atmospheric pathlength of a straight line (towards Earth) from a given position on Titan's observed disk is enlarged with respect to the minimum atmospheric pathlength, at the sub-Earth point in Titan - where $A(\theta = 0) = 1$. This airmass is dependent on the polar angle $\theta$ on Titan's sphere, and is given is given by eq. \ref{Airmass} following the plane-parallel atmosphere approximation.

\begin{equation}
    A(\theta) = \frac{1}{\cos(\theta)}
    \label{Airmass}
\end{equation}

We wish to calculate the integrated airmass $<A>$ over the disk of Titan observed by us. This requires an integration of airmass $A(\theta)$ over the whole of the observed apparent disk of Titan. Considering we observed Titan centred on its apparent disk's centre, we shall integrate over the polar angle  between the centre of the disk, $\theta = 0$, up until the maximum observed polar angle, $\theta = \alpha$, which is defined by the relative size of VLT-ESPRESSO's fiber (0.5'') with respect to Titan's apparent size (0.826''), on the night of the observation, as 

\begin{equation}
    \alpha = \arcsin\left( \frac{0.5''}{0.826''}\right) \approx 37.25º
    \label{alpha}
\end{equation}

Hence, the integration of $A(\theta)$ over the apparent observed disk of Titan, (normalized by the integrated element of apparent surface area) is given by

\begin{equation}
    <A> = \frac{\int \int A(\theta) dS_{APP}}{\int\int dS_{APP}} = \frac{\int_0^{2\pi} d\phi \int_0^\alpha A(\theta)R^2\cos^2(\theta)\sin(\theta) d\theta}{\int_0^{2\pi} d\phi\int_0^\alpha R^2\cos^2(\theta)\sin(\theta) d\theta}
\end{equation}

\par Replacing $A(\theta)$ by eq. \ref{Airmass}, we get

\begin{equation}
    <A> = \frac{\int_0^\alpha \cos(\theta)\sin(\theta) d\theta}{\int_0^\alpha \cos^2(\theta) \sin(\theta) d\theta} = \frac{3}{2}\cdot\frac{1 - \cos^2(\alpha)}{1 - \cos^3(\alpha)}
\end{equation}

We must consider a 2-way airmass for Titan, given that the radiation interacts twice with the column of C$_3$ absorbing gas: when it arrives to Titan, descending all the way to Titan's optical radius, and after being backscattered upwards, close to Titan's optical radius at altitudes of 300km (for 400nm photons), on its way out of Titan's atmosphere. Hence the integrated 2-way airmass on the observed disk of Titan, $\tilde{A}$, is given by eq. \ref{2-way-airmass}.

\begin{equation}
    \tilde{A} = 3\cdot\frac{1 - \cos^2(\alpha)}{1 - \cos^3(\alpha)}
    \label{2-way-airmass}
\end{equation}

\par If we had observed the entirety of Titan's disk, $\alpha = \pi/2$, which would yield an integrated 2-way airmass of 3. However, the incomplete coverage of Titan's disk by the VLT-ESPRESSO fiber prevents a stronger contribution to the airmass from Titan's limb. This yields, for $\alpha$ given by eq. \ref{alpha}, an integrated 2-way airmass on Titan of $\tilde{A} \simeq 2.2155$, for this observation.

\par To assess whether this plane-parallel atmosphere approximation is valid in the case of such an extended atmosphere as Titan's, we check how different the maximum observed polar angle, $\alpha'$, would be in a spherical atmosphere at the altitude at the peak of C$_3$ absorption (expected to occur below altitudes of 500km, \cite{Dobrijevic2016} \cite{Rianço-Silva2024}), when the plane-parallel $\alpha$ was obtained for the optical radius altitude (300km). This corrected maximum observed polar angle $\alpha'$ is obtained by eq. \ref{alpha_line}:

\begin{equation}
    \alpha' = \arccos\left[
        \sqrt{1 - \left( \frac{R}{R+H} \sin(\alpha) \right)^2 }
    \right]
    \label{alpha_line}
\end{equation}

with $R$ as the optical radius of Titan (equal to the solid body of Titan + 300 km at $\lambda=400$nm, \citep{Lorenz1999}) and $H$ as the altitude of the layer of interest above R (i.e., the expected altitude of maximum C$_3$ absorption at 500 km, and hence H = 200 km above the optical radius of Titan), and $\alpha$ as the maximum observed polar angle at the optical radius, given by eq. \ref{alpha}. Using this, we retrieve $\alpha'$ = 34.36º. Applying this new angle to the integrated airmass formula (eq.\ref{2-way-airmass}), we get $\tilde{A'}$ = 2.1845 which compares to the $\tilde{A}$ = 2.2155 integrated airmass obtained from the plane-parallel approximation, deviating by less than 1.5\% with respect to one another.

\par One other example of the effect of spherical geometry on an atmosphere which is not accounted by the plane-parallel approximation is the correct expression for the Airmass at a given zenithal angle. The plane-parallel atmosphere approximation describes Airmass by eq.\ref{Airmass}, which diverges for $\theta = 90º$. A commonly used spherical correction for this effect is given in \citep{KastenYoung1989}. Using their formulation, we numerically integrated this Airmass over an apparent planetary disk, up to a maximum angle $\alpha$. We noticed that for maximum integration angles (maximum observed polar angles) below 40º, the plane-parallel and the spherical geometry formulation of \citep{KastenYoung1989} do not differ by more than 0.1\%. Taking this into account, we conclude that the plane-parallel approximation provides a maximum deviation from spherical geometry integrated airmass “true” value of less than 1.5\%, and hence constitutes a valid approximation for this observation, when only the central regions of Titan's apparent disk are probed.

\section{Spectral Residuals}
\label{Appendix_Residuals}
We display here the residual plots resulting from the fit of two spectral models, one without C$_3$ and another corresponding to the MCMC best fit to the data, with respect to the VLT-ESPRESSO data of Titan. This is a direct comparison with figure \ref{fig:best_fit} where spectral models are compared to the observed spectrum. In these residuals, we subtract the models to the observed spectrum, assessing how different are the residuals for the no-C$_3$ case with respect to the MCMC retrieval best fit model which requires the presence of C$_3$ as described in section \ref{MCMC_section}. The shaded gray area corresponds to the region where residuals are smaller than the observed spectrum's uncertainty, which means that the model agrees with the observation in those wavelengths.

\begin{figure}
    \centering
    \includegraphics[width=\linewidth]{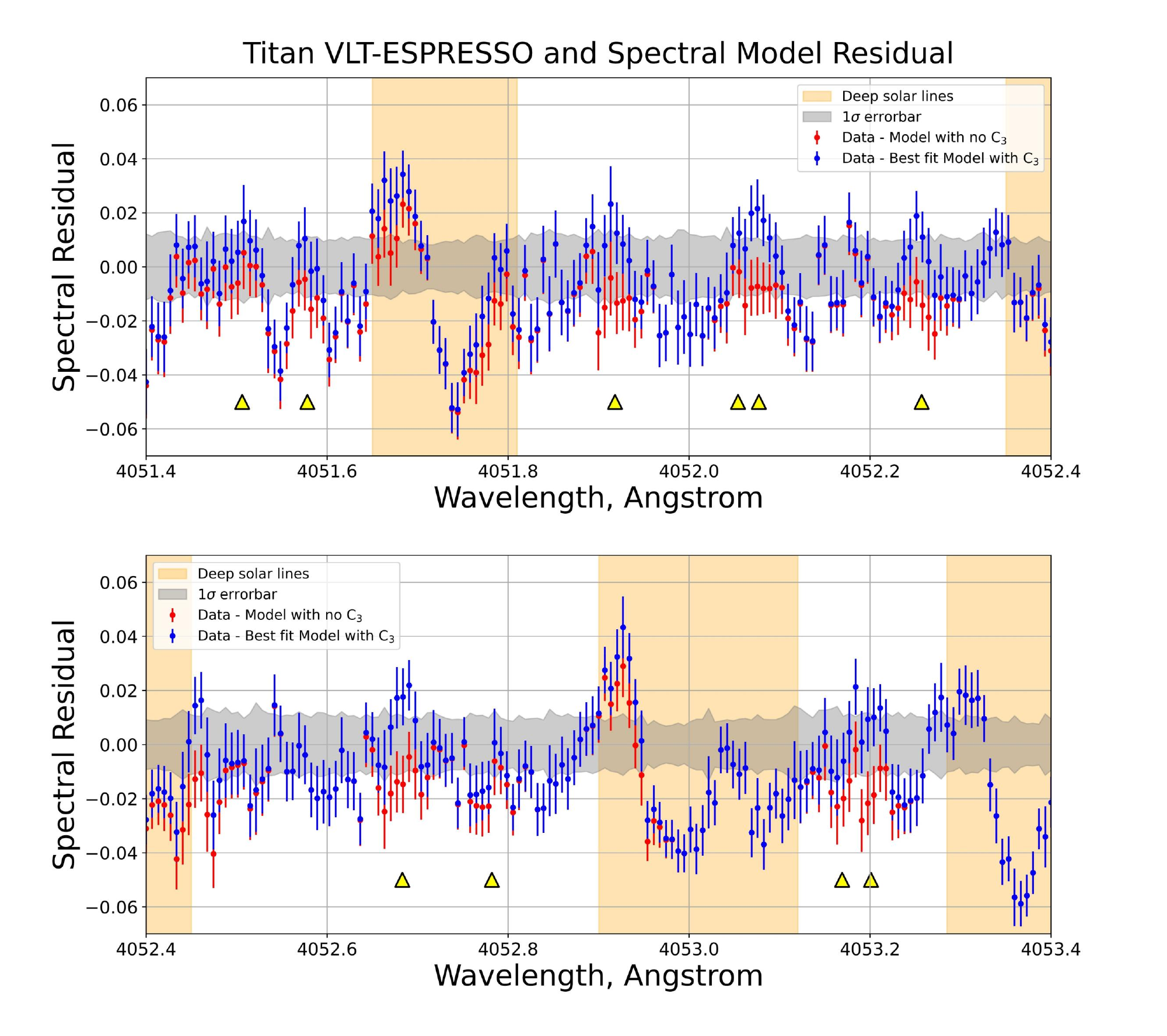}
    \caption{Residual plots of two distinct spectral models with respect to VLT-ESPRESSO spectrum of Titan. This figure directly compares with figure \ref{fig:best_fit}, where in red we showcase the residuals resulting from the subtraction of no-C$_3$ spectral model to the observed spectrum of Titan, and in blue we showcase the residuals resulting from the subtraction of the MCMC-best fit model (with C$_3$) to the observed spectrum of Titan. Shaded gray area correspond to the region around residuals = 0 smaller than the Titan's observed spectrum 1$\sigma$ errorbars - i.e., region of model agreement with observation within its uncertainty.The shaded orange region corresponds to spectral regions dominated by deep solar absorption lines (> 30\% of relative depth). Yellow markers show the central wavelength of the C$_3$ spectral lines according to \citep{Fan2024}.}
    \label{fig:Original_LL_residuals}
\end{figure}

\par As expected, we observe that the spectral regions where the two residuals differ align with the C$_3$ modelled spectral lines. Interestingly, in these regions, the residual of the MCMC best fit model (with C$_3$) is overall closer or within the shaded region that corresponds to the observations - i.e., as shown above with the $\chi^2$ analysis and the MCMC fit, the spectral model containing C$_3$ provides a better fit to the observed spectrum of Titan than the spectral model without C$_3$.

\section{C$_3$ Linelist and Line detection}
\label{Appendix_C3} We display here table \ref{tab:linelist_c3} containing the linelist of C$_3$ used to model the absorption spectrum of C$_3$ in this work. It also includes the detected line wavelengths, line depth with respect to the solar spectrum and detection significance (ratio between line depth and VLT-ESPRESSO errorbars at line center).

\onecolumn

\onecolumn

\begin{longtable}{|c|c|c|c|c|c|c|}
\caption{Linelist for the C$_3$ $\tilde{A} - \tilde{X}$ 000-000 band based on \protect\cite{Fan2024}, showcasing each identified transition (P(J), Q(J), R(J)), original wavelength $\lambda_j$ in \AA, oscillator strength $f_j$, and the observed matching features in wavelength, observed relative line depth, and detection significance. “O.D.S.L.” indicates overlap with a deep solar line (depth $>$10\%), preventing identification. ** marks C$_3$ lines that may be merged with nearby transitions. The last column shows the wavelength shifts used in the adapted linelist test in Appendix~\ref{Appendix_Adapted_Linelist}.}
\label{tab:linelist_c3}
\\
\hline
\shortstack{Rotational\\Line} &
\shortstack{Original\\$\lambda_j$ (\AA)} &
\shortstack{$f_j$\\(x$10^3$)} &
\shortstack{Observed\\$\lambda_j$ (\AA)} &
\shortstack{Observed\\Line Depth (\%)} &
\shortstack{Line\\Significance} &
\shortstack{Adapted Linelist\\Shift $\Delta\lambda_j$ (\AA)}
\\ \hline
\endfirsthead

\hline
\shortstack{Rotational\\Line} &
\shortstack{Original\\$\lambda_j$ (\AA)} &
\shortstack{$f_j$\\(x$10^3$)} &
\shortstack{Observed\\$\lambda_j$ (\AA)} &
\shortstack{Observed\\Line Depth (\%)} &
\shortstack{Line\\Significance} &
\shortstack{Adapted Linelist\\Shift $\Delta\lambda_j$ (\AA)}
\\ \hline
\endhead

\hline
\endfoot

\hline
\endlastfoot

R(20)                                                      & 4049.808                                                                                         & 4.29                                                                      & O.D.S.L.                                                                                           & -                                                                   & -                                                                     & -                                                                                                              \\ \hline
R(18)                                                      & 4049.861                                                                                         & 4.32                                                                      & O.D.S.L.                                                                                           & -                                                                   & -                                                                     & -                                                                                                              \\ \hline
R(16)                                                      & 4049.962                                                                                         & 4.36                                                                      & O.D.S.L.                                                                                           & -                                                                   & -                                                                     & -                                                                                                              \\ \hline
R(14)                                                      & 4050.075                                                                                         & 3.49                                                                      & O.D.S.L.                                                                                           & -                                                                   & -                                                                     & -                                                                                                              \\ \hline
R(12)                                                      & 4050.191                                                                                         & 3.58                                                                      & O.D.S.L.                                                                                           & -                                                                   & -                                                                     & -                                                                                                              \\ \hline
R(10)                                                      & 4050.327                                                                                         & 3.73                                                                      & O.D.S.L.                                                                                           & -                                                                   & -                                                                     & -                                                                                                              \\ \hline
R(6)*                                                      & 4050.401                                                                                         & 0.58                                                                      & O.D.S.L.                                                                                           & -                                                                   & -                                                                     & -                                                                                                              \\ \hline
R(8)                                                       & 4050.484                                                                                         & 3.95                                                                      & O.D.S.L.                                                                                           & -                                                                   & -                                                                     & -                                                                                                              \\ \hline
R(4)*                                                      & 4050.567                                                                                         & 0.49                                                                      & O.D.S.L.                                                                                           & -                                                                   & -                                                                     & -                                                                                                              \\ \hline
R(6)                                                       & 4050.661                                                                                         & 4.21                                                                      & O.D.S.L.                                                                                           & -                                                                   & -                                                                     & -                                                                                                              \\ \hline
Q(2)*                                                      & 4050.746                                                                                         & 0.45                                                                      & O.D.S.L.                                                                                           & -                                                                   & -                                                                     & -                                                                                                              \\ \hline
R(4)                                                       & 4050.857                                                                                         & 4.44                                                                      & O.D.S.L.                                                                                           & -                                                                   & -                                                                     & -                                                                                                              \\ \hline
R(2)                                                       & 4051.055                                                                                         & 3.78                                                                      & O.D.S.L.                                                                                           & -                                                                   & -                                                                     & -                                                                                                              \\ \hline
R(2)*                                                      & 4051.190                                                                                         & 2.08                                                                      & O.D.S.L.                                                                                           & -                                                                   & -                                                                     & -                                                                                                              \\ \hline
R(0)                                                       & 4051.255                                                                                         & 4.22                                                                      & O.D.S.L.                                                                                           & -                                                                   & -                                                                     & -                                                                                                              \\ \hline
R(0)*                                                      & 4051.396                                                                                         & 10.58                                                                     & O.D.S.L.                                                                                           & -                                                                   & -                                                                     & -                                                                                                              \\ \hline
Q(2)                                                       & 4051.448                                                                                         & 6.87                                                                      & O.D.S.L.                                                                                           & -                                                                   & -                                                                     & -                                                                                                              \\ \hline
Q(4)                                                       & 4051.506                                                                                         & 7.70                                                                      & 4051.54                                                                                            & 4.3                                                                 & 3.2 $\sigma$                                            & + 0.04                                                                                                         \\ \hline
Q(6)                                                       & 4051.578                                                                                         & 7.89                                                                      & 4051.60                                                                                            & 3.1                                                                 & 2.6 $\sigma$                                            & +0.03                                                                                                          \\ \hline
Q(8)                                                       & 4051.670                                                                                         & 7.95                                                                      & O.D.S.L.                                                                                           & -                                                                   & -                                                                     & -                                                                                                              \\ \hline
Q(10)                                                      & 4051.782                                                                                         & 7.96                                                                      & O.D.S.L.                                                                                           & -                                                                   & -                                                                     & -                                                                                                              \\ \hline
P(2)                                                       & 4051.820                                                                                         & 1.06                                                                      & O.D.S.L.                                                                                           & -                                                                   & -                                                                     & -                                                                                                              \\ \hline
Q(12)                                                      & 4051.918                                                                                         & 7.97                                                                      & 4051.90                                                                                            & 2.7                                                                 & 2.1 $\sigma$                                            & -0.01                                                                                                          \\ \hline
P(4)                                                       & 4052.054                                                                                         & 1.57                                                                      & 4052.02                                                                                            & 2.4                                                                 & 2.2 $\sigma$                                            & -0.03                                                                                                          \\ \hline
Q(14)                                                      & 4052.077                                                                                         & 7.98                                                                      & 4052.11**                                                                                          & 2.0                                                                 & 2.4 $\sigma$                                            & +0.04                                                                                                          \\ \hline
Q(6)*                                                      & 4052.122                                                                                         & 0.28                                                                      & No Match**                                                                                         & -                                                                   & -                                                                     & -                                                                                                              \\ \hline
P(4)*                                                      & 4052.180                                                                                         & 0.86                                                                      & No Match**                                                                                         & -                                                                   & -                                                                     & -                                                                                                              \\ \hline
Q(16)                                                      & 4052.257                                                                                         & 7.98                                                                      & 4052.27                                                                                            & 1.9                                                                 & 1.6 $\sigma$                                            & +0.01                                                                                                          \\ \hline
P(6)                                                       & 4052.412                                                                                         & 2.56                                                                      & O.D.S.L.                                                                                           & -                                                                   & -                                                                     & -                                                                                                              \\ \hline
Q(18)                                                      & 4052.459                                                                                         & 7.99                                                                      & O.D.S.L.                                                                                           & -                                                                   & -                                                                     & -                                                                                                              \\ \hline
P(8)*                                                      & 4052.521                                                                                         & 0.39                                                                      & O.D.S.L.                                                                                           & -                                                                   & -                                                                     & -                                                                                                              \\ \hline
Q(20)                                                      & 4052.683                                                                                         & 7.99                                                                      & 4052.67                                                                                            & 3.2                                                                 & 2.2 $\sigma$                                            & -0.01                                                                                                          \\ \hline
P(8)                                                       & 4052.782                                                                                         & 2.82                                                                      & 4052.77                                                                                            & 2.3                                                                 & 2.4 $\sigma$                                            & -                                                                                                              \\ \hline
Q(22)                                                      & 4052.938                                                                                         & 7.99                                                                      & O.D.S.L.                                                                                           & -                                                                   & -                                                                     & -                                                                                                              \\ \hline
P(10)                                                      & 4053.169                                                                                         & 2.87                                                                      & 4053.15                                                                                            & 2.3                                                                 & 1.8 $\sigma$                                            & -                                                                                                              \\ \hline
Q(24)                                                      & 4053.201                                                                                         & 7.99                                                                      & 4053.19                                                                                            & 3.5                                                                 & 2.7 $\sigma$                                             & -                                                                                                              \\ \hline
Q(26)                                                      & 4053.479                                                                                         & 7.99                                                                      & O.D.S.L.                                                                                           & -                                                                   & -                                                                     & -                                                                                                              \\ \hline
P(12)                                                      & 4053.577                                                                                         & 2.87                                                                      & O.D.S.L.                                                                                           & -                                                                   & -                                                                     & -                                                                                                              \\ \hline
Q(28)                                                      & 4053.783                                                                                         & 8.00                                                                      & O.D.S.L.                                                                                           & -                                                                   & -                                                                     & -                                                                                                              \\ \hline
P(14)                                                      & 4054.005                                                                                         & 2.86                                                                      & O.D.S.L.                                                                                           & -                                                                   & -                                                                     & -                                                                                                              \\ \hline
Q(30)                                                      & 4054.108                                                                                         & 8.00                                                                      & O.D.S.L.                                                                                           & -                                                                   & -                                                                     & -                                                                                                              \\ \hline
P(16)                                                      & 4054.447                                                                                         & 3.64                                                                      & O.D.S.L.                                                                                           & -                                                                   & -                                                                     & -                                                                                                              \\ \hline
P(18)                                                      & 4054.902                                                                                         & 3.68                                                                      & O.D.S.L.                                                                                           & -                                                                   & -                                                                     & -                                                                                                              \\ \hline
P(20)                                                      & 4055.369                                                                                         & 3.71                                                                      & O.D.S.L.                                                                                           & -                                                                   & -                                                                     & -                                                                                                              \\ \hline

\end{longtable}

\twocolumn

\section{Adapted C$_3$ Linelist and retrieval}
\label{Appendix_Adapted_Linelist}

\begin{figure}	\includegraphics[width=\linewidth]{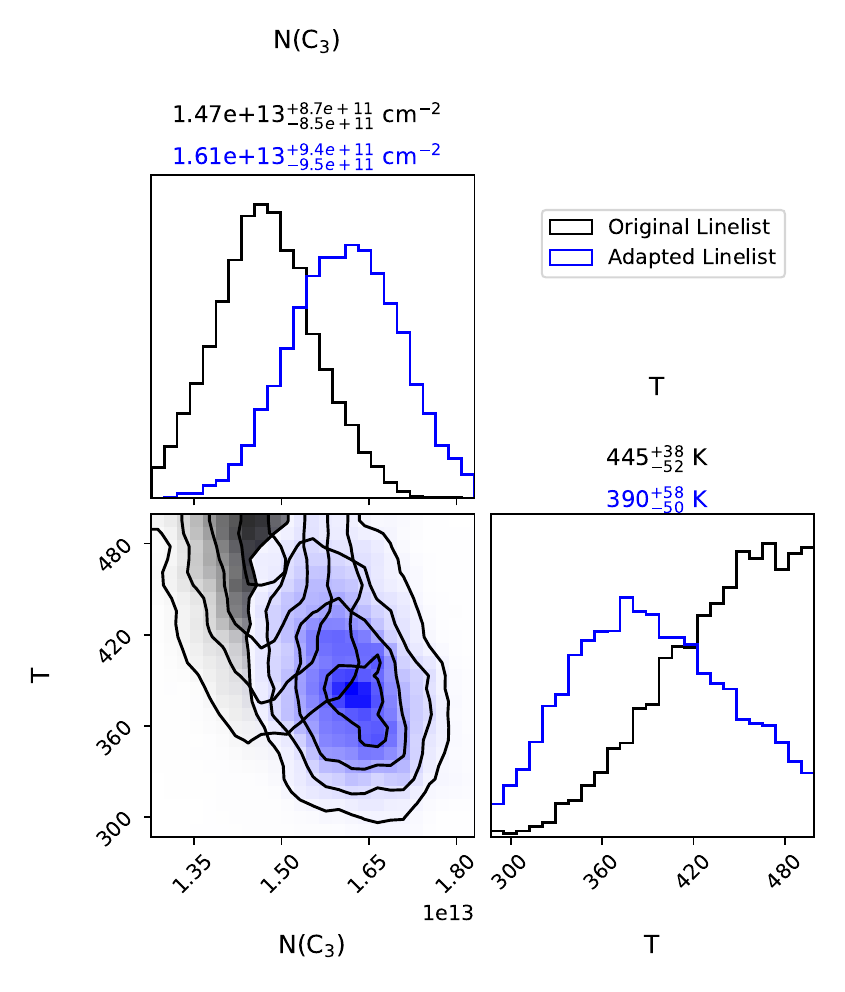}
    \caption{Posterior plots from the MCMC fit to the VLT-ESPRESSO spectrum of Titan - for the original \citep{Fan2024} C$_3$ (black) and our suggested adapted linelist, which improves the spectral model match to the observed spectral features (blue). Fit for the column density of C$_3$, ($N$) and for the Temperature ($T$). Retrieved values showcased with 1$\sigma$ errorbars. Contour plots on the 2D histogram correspond to the (0.5, 1, 1.5, 2)-sigma confidence regions for a 2D Gaussian distribution.}
    \label{fig:posteriors_original_vs_adapted}

\end{figure}

\begin{figure}	\includegraphics[width=\linewidth]{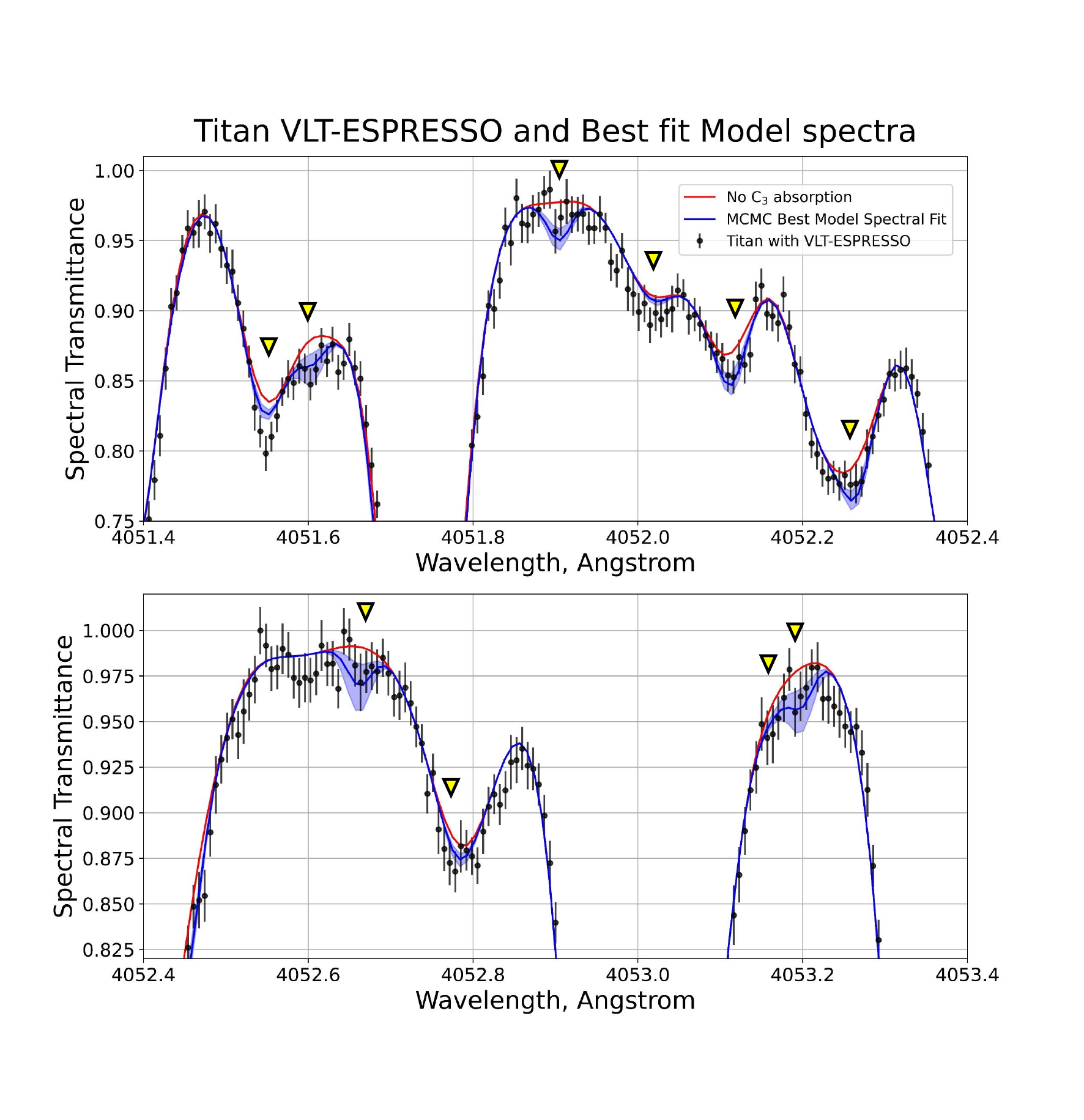}
    \caption{VLT-ESPRESSO normalized spectrum of Titan (black datapoints) at the spectral region of interest for the $\tilde{A}^1 \Pi_u - \tilde{X}^1\Sigma^+_g$ 000-000 band of C$_3$ compared to the backscattered solar spectrum (a proxy for Titan's spectrum with no C$_3$ absorption, in red) and to the best-fit model obtained from the MCMC likelihood fit with the adapted C$_3$ linelist described in \ref{tab:linelist_c3}, associated with the retrieved values for $N(C_3)$ and $T$, in blue. The shaded blue area corresponds to the best fit model uncertainty based on the retrieved $N(C_3)$ and $T$ values with 3$\sigma$ errorbars. The yellow triangles mark the identified spectral features we associate to C$_3$ absorption features, described in table \ref{tab:linelist_c3}.}
    \label{fig:line_detection_adapted_LL}

\end{figure}

\begin{figure}
    \centering
    \includegraphics[width=\linewidth]{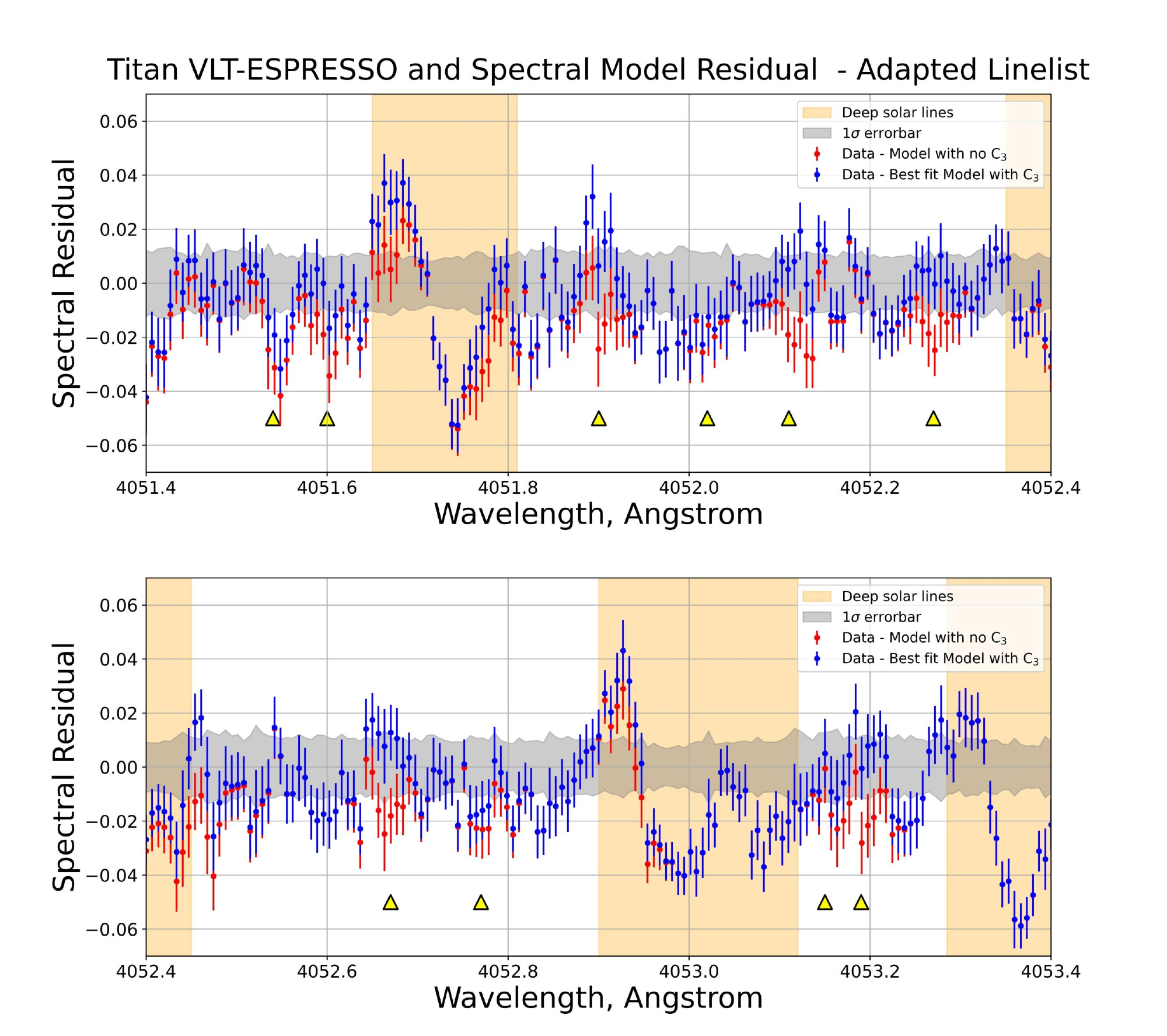}
    \caption{Residual plots of two distinct spectral models with respect to VLT-ESPRESSO spectrum of Titan. This figure directly compares with figure \ref{fig:Original_LL_residuals}, albeit with the adapted C$_3$ linelist, rather than the original linelist by \citep{Fan2024}. In red we showcase the residuals resulting from the subtraction of no-C$_3$ spectral model to the observed spectrum of Titan, and in blue we showcase the residuals resulting from the subtraction of the MCMC-best fit model (with C$_3$) to the observed spectrum of Titan. Shaded gray area correspond to the region around residuals = 0 smaller than the Titan's observed spectrum 1$\sigma$ errorbars - i.e., region of model agreement with observation within its uncertainty. The shaded orange region corresponds to spectral regions dominated by deep solar absorption lines (> 30\% of relative depth). Yellow markers show the central wavelength of the C$_3$ spectral lines according our adapted linelist.}
    \label{fig:Adapted_LL_residuals}

\end{figure}

MCMC retrievals provide an interesting way to compare between distinct models fits to the same dataset \citep{Waldmann2015}. After noticing a slight mismatch between some identified spectral features on the VLT-ESPRESSO spectrum of Titan and the expected C$_3$ line positions, we suggested these mismatches could be due to the empirical nature of the \citep{Fan2024} linelist, which was obtained at a worse spectral resolution than these VLT-ESPRESSO observations. Hence, at table \ref{tab:linelist_c3} we suggest small shifts (below the \citep{Fan2024} typical line wavelength uncertainty of 0.05\AA) that could be expected to improve the overall fit to the data. This was tested by applying the same MCMC retrieval test described in section \ref{sec:mcmc}, comparing the posteriors obtained by using the original \citep{Fan2024} versus the adapted linelists - as shown in figure \ref{fig:posteriors_original_vs_adapted}, with the best fit plot shown in figure \ref{fig:line_detection_adapted_LL} and the residual plots in figure \ref{fig:Adapted_LL_residuals}.

\par As expected, the adapted linelists corner plot and histograms show a better Temperature constraint (in blue, following a distribution closer to a Gaussian curve within the allowed prior range) than the original linelist (black). The retrieved C$_3$ abundance using the adapted linelist is shown (from its marginalized histogram) to be slightly higher for the original linelist, at $N(C_3)$, with $N(C_3) = \left(1.61\pm 0.09\right) \times 10^{13}$ cm$^{-2}$ at 1$\sigma$, $N(C_3) = \left(1.61^{+\hspace{0.5mm}0.29}_{-\hspace{0.5mm}0.28}\right) \times 10^{13}$ cm$^{-2}$ at 3$\sigma$ and $N(C_3) = \left(1.61^{+\hspace{0.5mm}0.34}_{-\hspace{0.5mm}0.36} \right) \times 10^{13}$ cm$^{-2}$ at 5$\sigma$ - still within the expected errorbars obtained with the original linelist. Retrieved temperature values are nonetheless different, due to the better constraint this alternative linelist enabled: $T = (390^{+58}_{-50})K$ at 1$\sigma$, $T = (390^{+110}_{-126})K$ at 3$\sigma$, and $T = (390^{+110}_{-157})K$ at 5$\sigma$. A possible explanation is that due to the enhanced fit the spectrum obtained with the adapted linelist, the retrieval is not as skewed towards higher temperatures, which tend to improve this fit by broadening the modelled spectrum, bringing them closer to the slightly shifted spectral features in the data.

\par After confirming the detection of C$_3$ in Titan's atmosphere through the $\chi^2$ method (section \ref{section:chi2}) combined with a MCMC retrieval using the original \citep{Fan2024} linelist, the retrieval comparison shown could indicate that the adapted linelist may indeed be preferred when modelling the 405-nm band of C$_3$ at ultra-high-resolutions.

\bsp	
\label{lastpage}
\end{document}